\documentclass[useAMS,usenatbib,usegraphicx]{mn2e}

\usepackage{graphicx}
\usepackage[latin1]{inputenc}
\usepackage{color}
\usepackage{times}
\usepackage{natbib}
\usepackage{setspace}
\newif\ifAMStwofonts
\AMStwofontstrue
\definecolor{red}{rgb}{1,0.,0.}

\newcommand{\mor}{{\sc morgana}}
\newcommand{\mun}{WDL08}
\newcommand{\som}{S08}
\newcommand{\msun}{{\rm M}_\odot}

\def\lesssim{\lower.5ex\hbox{$\; \buildrel < \over \sim \;$}}
\def\gtrsim{\lower.5ex\hbox{$\; \buildrel > \over \sim \;$}}

\title[Downsizing in Hierarchical Models] {The Many Manifestations of
  Downsizing: Hierarchical Galaxy Formation Models confront
  Observations.}

\author[Fontanot et al.]{
  \parbox[t]{\textwidth}{ 
    Fabio Fontanot$^1$, 
    Gabriella De Lucia$^{2,3}$, 
    Pierluigi Monaco$^{3,4}$, 
    Rachel S. Somerville$^{1,5}$,
    Paola Santini$^{6,7}$ \\
    }
    \vspace*{6pt}\\
    $^1$ MPIA Max-Planck-Institute f\"ur Astronomie, Koenigstuhl 17, 69117 Heidelberg, Germany\\
    $^2$ MPA Max-Planck-Institute f\"ur Astrophysik, Karl-Schwarzschild-Strasse 1, D-85748, Garching, Germany \\
    $^3$ INAF-Osservatorio Astronomico di Trieste, Via Tiepolo 11, I-34131 Trieste, Italy \\
    $^4$ Dipartimento di Astronomia, Universit\`a di Trieste, via Tiepolo 11, 34131 Trieste, Italy \\
    $^5$ Space Telescope Science Institute, 3700 San Martin Drive, Baltimore, MD 21218, USA \\
    $^6$ INAF-Osservatorio Astronomico di Roma, via Frascati 33, I-00040 Monteporzio, Italy \\
    $^7$ Dipartimento di Fisica, Universit\`{a} di Roma ``La Sapienza'', P.le A. Moro 2, 00185 Roma\\
    email: fontanot@mpia.de, delucia@oats.inaf.it, monaco@oats.inaf.it, somer@stsci.edu, santini@mporzio.astro.it}

\begin{document}
\date{Accepted ... Received ...}

\maketitle

\begin{abstract} 
  It has been widely claimed that several lines of observational
  evidence point towards a ``downsizing'' of the process of galaxy
  formation over cosmic time. This behaviour is sometimes termed
  ``anti-hierarchical'', and contrasted with the ``bottom-up'' (small
  objects form first) assembly of the dark matter structures in Cold
  Dark Matter models. In this paper we address three different kinds
  of observational evidence that have been described as
  ``downsizing'': the stellar mass assembly (i.e.  more massive
  galaxies assemble at higher redshift with respect to low-mass ones),
  star formation rate (i.e. the decline of the specific star formation
  rate is faster for more massive systems) and the ages of the stellar
  populations in local galaxies (i.e. more massive galaxies host older
  stellar populations).  We compare a broad compilation of available
  data-sets with the predictions of three different semi-analytic
  models of galaxy formation within the $\Lambda$CDM framework. In the
  data, we see only weak evidence at best of ``downsizing'' in stellar
  mass and in star formation rate. Despite the different
  implementations of the physical recipes, the three models agree
  remarkably well in their predictions. We find that, when
  observational errors on stellar mass and SFR are taken into account,
  the models acceptably reproduce the evolution of massive galaxies
  ($M>10^{11}\ \msun$ in stellar mass), over the entire redshift range
  that we consider ($0 \lesssim z \lesssim 4$).  However, lower mass
  galaxies, in the stellar mass range $10^9-10^{11}\ \msun$, are
  formed too early in the models and are too passive at late times.
  Thus, the models do not correctly reproduce the downsizing trend in
  stellar mass or the archaeological downsizing, while they
  qualitatively reproduce the mass-dependent evolution of the SFR. We
  demonstrate that these discrepancies are not solely due to a poor
  treatment of satellite galaxies but are mainly connected to the
  excessively efficient formation of central galaxies in high-redshift
  haloes with circular velocities $\sim 100-200$ km/s.  We conclude
  that some physical process operating on these mass scales --- most
  probably star formation and/or supernova feedback --- is not yet
  properly treated in these models.
\end{abstract}


\section{Introduction}\label{intro}

In the last decades, the parameters of the cosmological model have
been tightly constrained \citep[][and references therein]{Komatsu09},
and the CDM paradigm has proved to be very successful in reproducing a
large number of observations, particularly on large scales. The
current standard paradigm for structure formation predicts that the
collapse of dark matter (DM) haloes proceeds in a ``bottom-up''
fashion, with smaller structures forming first and later merging into
larger systems. It has long been known that galaxies do not share the
same `bottom-up' evolution, at least in their star formation
histories. The most massive galaxies --- mainly giant ellipticals
hosted in galaxy groups and clusters --- are dominated by old stellar
populations. In contrast, faint field galaxies appear to have
continued to actively form stars over the last billion years, and
their stellar populations are dominated by young stars. This evidence
is not necessarily in contrast with the hierarchical clustering of DM
haloes as it relates to the ``formation'' of the main stellar
population of a galaxy, which does not necessarily coincide with the
``assembly'' of its stellar mass and/or the assembly of its parent
dark matter halo.

In the last decade, much observational effort has been devoted to
quantifying the dependence of galaxy formation and assembly on stellar
mass. In one of the earliest such studies, \citet{Cowie96} showed that
the maximum rest-frame $K$-band luminosity of galaxies undergoing
rapid star formation in the Hawaii deep field declines smoothly with
cosmological time.  Cowie and collaborators coined the term
``downsizing'' (hereafter DS) to describe this behaviour. Since then,
the same term has been extended to a number of observational trends
suggesting either older ages, earlier active star formation, or
earlier assembly for more massive galaxies with respect to their lower
mass counterparts. Using the same word to describe very different
kinds of observational results has naturally generated some confusion.
The underlying thought has clearly been that these
observations are all {\it manifestations} of the same underlying
physical process. It is not clear to which degree this is in fact the
case, nor to what degree these observational trends are
``anti-hierarchical'', i.e. whether they are in fact in serious
conflict with predictions from models based on $\Lambda$CDM cosmology.

It is useful, at this point, to summarise the different types of
``downsizing'' that have been discussed in the literature. Clearly,
each of the observational evidences discussed below has its own set of
uncertainties and potential biases.  Here, we report the trends as
they have been claimed in the literature, and discuss in more detail
the related uncertainties and caveats later. The first two types of DS
that we describe are based on the local ``fossil record'' and are
related to the time of ``formation'' of the stellar population,
i.e. they tell us that the bulk of the stars in more massive galaxies
formed earlier and on shorter time-scales than in their lower mass
counterparts. These two types of DS are:

{\em Chemo-archaeological DS}: among elliptical galaxies, more
massive objects have higher (up to super-solar) [$\alpha$/Fe] ratios.
This result was first reported by \citet{Faber92} and
\citet{Worthey92}, who suggested three possible (and equally
acceptable at that time) explanations: (i) different star formation
time-scales; (ii) a variable Initial Mass Function (IMF); (iii)
selective mass loss mechanisms. Several studies have since confirmed
this observational trend \citep{Carollo93, Davies93, Trager00a,
  Kuntschner01}, and a standard interpretation has become that of
shorter formation time-scales in more luminous/massive galaxies
\citep{Matteucci94,Thomas05}, though other interpretations have not
been conclusively ruled out.

{\em Archaeological DS}: more massive galaxies host older stellar
populations than lower mass galaxies.  A direct estimate of stellar
ages is hampered by the well-known age-metallicity degeneracy
\citep[e.g.][and references therein]{Trager00b}, although it has long
been known that there are some spectral features (like Balmer lines)
that are more sensitive to age than to metallicity \citep[see
  i.e.,][]{Worthey94}. Recent detailed analyses, based on a
combination of spectral indexes or on a detailed fit of the full
high-resolution spectrum, have confirmed a weak trend between stellar
mass and age both in clusters \citep[][though see Trager et
  al. 2008]{Thomas05,Nelan05}, and in the field
\citep{Trager00b,Heavens04, Gallazzi05, Panter07}.

The second kind of observational evidence for DS comes from ``lookback
studies'', or observations of galaxies at different cosmic
epochs: 

{\em DS in (specific) SFR}: the mass of `star-forming galaxies'
declines with decreasing redshift.  This trend was first seen by
\citet{Cowie96}, and there have been many claimed confirmations by
subsequent deeper and/or wider observational programmes
\citep{Brinchmann04, Kodama04, Feulner05, Bauer05, Papovich06,
  Bundy06, Pannella06, Bell07, Noeske07, Drory08, Vergani08, Chen09,
  Cowie08}. This trend can also be recast as implying that the SFR
density or specific star formation rate declines more rapidly for more
massive systems; here there are conflicting claims in the literature
about whether such a trend is in fact seen or not
\citep[e.g.][]{Juneau05,Zheng07,Conselice07,Mobasher09}. This trend
reflects in SSFRs of nearby spiral galaxies which are higher for lower
mass objects \citep{Boselli01}.  A possibly related trend is the
increase with time of faint red-sequence galaxies in galaxy clusters
\citep[see, e.g.,][]{DeLucia04c, DeLucia07a, Gilbank08}, which may be
due to a differential decline of the SSFR.

{\em DS in stellar mass}: the high-mass end of the stellar mass
function (hereafter MF) evolves more slowly than the low-mass end,
indicating that massive galaxies were assembled earlier than less
massive ones. The same result is found both by correcting the B-band
or K-band luminosity functions for ``passive'' evolution
\citep{Cimatti06} and by estimating the stellar mass using
multi-wavelength photometry \citep{Drory04, Drory05, Bundy06, Borch06,
  Fontana06, Pozzetti07, Conselice07, PerezGonzalez08,
  Marchesini08}. The significance of these claims has been recently
questioned by \citet{Marchesini08}.

{\em DS in metallicity}: the stellar metallicity of more massive
galaxies appears to decrease with redshift more slowly than for less
massive galaxies \citep{Savaglio05, Erb06, Ando07, Maiolino08}. It is
important to note, however, that often different indicators are
used at different redshifts, and that there are large uncertainties in the
metallicity calibration \citep{Kewley08}.

{\em DS in nuclear activity}: the number density of Active Galactic
Nuclei (AGNs) peaks at higher redshift when brighter objects are
considered.  This trend is found both for X-ray \citep{Ueda03,
  Hasinger05} and for optically \citep{Cristiani04, Fontanot07a}
selected AGNs, but it strongly depends on the modeling of obscuration
\citep[e.g.,][]{LaFranca05}.

Downsizing trends have often been considered ``anti-hierarchical'',
suggesting expected and/or demonstrated difficulties in reconciling
the observed trends with predictions from hierarchical galaxy
formation models. The naive expectation is that, like for dark matter
haloes, galaxy formation also proceeds in a bottom-up fashion with
more massive systems ``forming'' later.  It has already been pointed
out in early theoretical work \citep{BaughColeFrenk96,Kauffmann96}
that the epoch of formation of the stars within a galaxy does not
necessarily coincide with the epoch of the galaxy's
assembly. Moreover, \citet{Neistein06} \citep[see also][]{Li08}
suggested that a certain degree of ``natural downsizing'' is actually
expected in the CDM paradigm if one assumes that there is a minimum
halo mass that can support star formation and considers the integrated
mass in all progenitor haloes rather than just that in the main
progenitor.  However, several authors \citep{Cimatti06, Fontana06,
  Fontanot06, Cirasuolo08} have argued that the observed mass assembly
DS represents a challenge for modern hierarchical galaxy formation
models. As well, CDM models have been unable\footnote{It should be
  noted that the work by \citet{Thomas99} was not fully
  self-consistent as they used star formation histories from a
  semi-analytic model in a ``closed-box'' chemical enrichment model.
  The later work by \citet{Nagashima05} has, however, confirmed the
  difficulties in reproducing the observed trends.} to reproduce the
observed chemo-archaeological DS \citep{Thomas99, Thomas05,
  Nagashima05, Pipino08}, and \citet{Somerville08} and
\citet{TragerSomerville08} have shown that the modern generation of
models does not quantitatively reproduce the archaeological DS trend
in the field or in rich clusters.

Early phenomenological models of joint galaxy-AGN formation by
\citet{Monaco00} and \citet{Granato01} produced "anti-hierarchical"
formation of elliptical galaxies in $\Lambda$CDM halos by delaying
quasar activity in less massive halos. More recently, it has been
suggested that AGN feedback could provide a solution to the
``downsizing problem'' \citep{Bower06,Croton06}. The suppression of
late gas condensation in massive haloes gives rise to shorter
formation time-scales for more massive galaxies \citep{DeLucia06}, in
qualitative agreement with the observed trends. However, the recent
work by \citet{Somerville08} indicates that the predicted trends may
not be as strong as the observed ones, even in the presence of AGN
feedback.

Moreover, AGN feedback does not stop the growth in stellar mass via
mergers. $\Lambda$CDM models predict that the stellar masses of the
most massive galaxies have increased by a factor two or more since
$z\sim1$ via gas-poor ``dry mergers'' \citep{DeLucia06,DeLucia07b}. It
has been suggested that if mergers scatter a significant fraction of
the stars in the progenitor galaxies into a ``diffuse stellar
component'', then perhaps one can reconcile the CDM
predictions with the observed weak evolution in the stellar mass
function since $z\sim 1$ \citep{Monaco06, Conroy07, Somerville08}, but
observational uncertainties on the amount of diffuse light are still
too large to strongly constrain models of this process.

Despite the large number of papers related to the subject of
``downsizing'', a detailed and systematic comparison between a broad
compilation of observational data and predictions from hierarchical
galaxy formation models is still missing. Our study is a first attempt
in this direction. We present here predictions from three different
semi-analytic models (see \S\ref{sec:models}), all of which have been
tuned to provide reasonably good agreement with the observed
properties of galaxies in the local Universe, and compare them to an
extensive compilation of recent data on the evolution of the stellar
MF, SSFRs, and SFR densities, as well as with observational
determinations of stellar population ages as a function of mass in
nearby galaxies. An important new aspect of our study is that we
consider three (claimed) ``manifestations'' of downsizing
simultaneously. Because these very different kinds of observations
have very different potential selection effects and biases, this
allows us to make a strong argument that, when a discrepancy is seen
between the models and all three kinds of observations, this
discrepancy is due to shortcomings in the physical ingredients of the
models rather than errors or biases in the observations. Similarly, by
making use of three independently developed semi-analytic models,
which include different implementations of the main physical
processes, we can hope to determine which conclusions are robust to
model details.

In this study, we do not address the ``Chemo-archaeological DS'' or
the ``DS in metallicity''. Our chemical enrichment models are all
based on an instantaneous recycling approximation which prevents us
from making detailed comparisons with observed elemental abundances.
We have also decided not to discuss here the ``DS in AGN activity'',
which depends strongly on the complicated and poorly understood
physics of accretion onto black holes and on its relation to star
formation activity \citep[see, e.g.,][]{Menci04,Menci08,Fontanot06}.

The paper is organized as follows: in \S\ref{sec:models} we give a
brief introduction to the models we use in our study. We then present
our results for the DS in stellar mass (\S\ref{sec:mass_down}), in SFR
(\S\ref{sec:sfr_down}), and on the archaeological DS
(\S\ref{sec:arch_down}).  In \S\ref{sec:final}, we discuss our results
and give our conclusions. Throughout this paper we assume a
cosmological model consistent with the WMAP3 results.

\section{Models}
\label{sec:models}

We consider predictions from three independently developed codes that
use semi-analytic modeling (SAM) techniques to simulate the formation
of galaxies within the $\Lambda$CDM cosmogony \citep[for a review on
  these techniques see][]{Baugh06}. In SAMs, the evolution of the
baryonic component of galaxies --- which are assumed to form when gas
condenses at the center of DM haloes --- is modeled using simple but
physically motivated analytic `recipes'.  The parameters entering
these analytic approximations of the various physical processes are
usually fixed by comparing model predictions to observational data of
local galaxies. Although the treatment of the physical processes is
necessarily simplified, this technique allows modelers to explore (at
least schematically) a broad range of processes that could not be
directly simulated simultaneously (e.g., accretion onto SMBH on sub-pc
scales within the framework of cosmological structure formation), and
to explore a wide parameter space.

Most of the various SAMs proposed in the literature are attempting to
model the same basic set of physical processes. When a comparison is
made of several SAMs with observations, one may focus on differences
between models, with the aim of understanding how the details of a
particular implementation influence the predictions of galaxy
properties.  Alternatively, one may concentrate on comparing the model
predictions with the observational data. In this case the focus shifts
to assessing whether the general framework, namely $\Lambda$CDM $+$
the set of physical processes implemented, gives a plausible
description of galaxy populations.

In this paper we take the second approach. We use three SAMs: (i) the
most recent implementation of the Munich model \citep{DeLucia07b} with
its generalization to the WMAP3 cosmology discussed in
\citet[][hereafter {\mun}]{Wang08}; (ii) the {\mor} model, presented
in \citet{Monaco07}, adapted to a WMAP3 cosmology, and with some minor
improvements which will be presented in \citet{LoFaro09}, (iii) the
fiducial model presented in \citet[][hereafter {\som}]{Somerville08},
which builds on the previous implementation discussed in
\citet{Somerville01}. All models adopt, for the results discussed in
this study, a \citet{Chabrier03} IMF.

In the following we briefly summarize the main physical ingredients of
SAMs, and then highlight the main differences between the
implementations of these ingredients in the three models used
here. For more details we refer to the original papers mentioned above
and to the references therein.

We first summarize the elements that are common to all three
models. The backbone of all three SAM's is a ``merger tree'', which
describes the formation history of dark matter haloes through mergers
and accretion. When a halo merges with a larger virialized halo, it
becomes a ``sub-halo'' and continues to orbit until it is either
tidally destroyed or merges with the central object. Gas cools and
condenses via atomic cooling, and forms a rotationally supported
disc. This cold disc gas becomes available for star formation, which
is modeled using simple empirical (Schmidt-Kennicutt-like)
recipes. Galaxy mergers trigger enhanced ``bursts'' of star
formation. After the Universe becomes reionized, gas infall is
suppressed in low-mass haloes ($\lesssim 30$--50 km/s) due to the
photoionizing background. Star formation deposits energy into the cold
gas, and may re-heat or expel this gas. The production of chemical
elements by type II supernovae is tracked using a simple instantaneous
recycling approximation with the effective yield taken as a free
parameter. All three codes also track the formation of supermassive
black holes, and differentiate between the so-called ``bright-mode''
(or ``quasar-mode'') which is associated with luminous AGNs, and
``radio-mode'' accretion which is related to efficient production of
radio jets. The ``bright-mode'' is associated with galaxy-galaxy
mergers ({\mun} and {\som}) or Eddington-limited accretion rates
({\mor}), while the ``radio-mode'' is associated with low accretion
rates (few percent of Eddington). All three models include ``radio
mode'' feedback (heating of the hot gas halo by giant radio
jets). {\mor} and {\som} also include galactic scale AGN-driven winds,
which can remove cold gas from galaxies.

All three models are coupled with stellar population synthesis models
and a treatment of dust absorption, and are capable of predicting
observable quantities like luminosities and colours in various bands.
However, modeling these additional ingredients (especially dust, as
recently shown by \citealt{Fontanot08}) introduces a large number of
additional uncertainties and degrees of freedom in both the
model-model and the model-data comparison. To simplify the
interpretation of our results, we therefore conduct our entire
analysis in the space of ``physical'' quantities (e.g. stellar masses
and SFRs), which are directly predicted by the models, and may be
extracted from multi-wavelength observations.

Here we highlight a few of the differences between the model
implementations:

\begin{itemize}

\item {\bf Cosmological and Numerical Parameters}: All three
  semi-analytic models adopt values of the cosmological parameters
  that are consistent with the WMAP3 results within the quoted
  errors. {\mun} use a simulation box of $125 h^{-1} {\rm Mpc}^3$ on a
  side, with cosmological parameters
  $(\Omega_0,\Omega_\Lambda,h,\sigma_8,n_{\rm sp})=(0.226, 0.774,
  0.743, 0.722, 0.947)$.  {\mor} uses a 144~$h^{-1}$~Mpc box, and
  adopts a cosmology with $(\Omega_0,\Omega_\Lambda,h,\sigma_8,n_{\rm
    sp})=(0.24, 0.76, 0.72, 0.8, 0.96)$. {\som} uses a grid of 100
  realizations of 100 ``root'' haloes, with circular velocities
  ranging from 60 km/s to 1200 km/s, and weights the results with the
  \citet{ShethTormen99} halo mass function. The {\som} model used here
  assumes the following cosmological parameters
  $(\Omega_0,\Omega_\Lambda,h,\sigma_8,n_{\rm sp})=(0.279, 0.721,
  0.701, 0.817, 0.96)$. In all cases, the mass resolution is
  sufficient to resolve galaxies with stellar mass larger than
  $10^9\,h^{-1}\ \msun$. The very small differences in the
  cosmological parameters in the three models will have a nearly
  undetectable impact on our predictions, and therefore we make no
  attempt to correct the results for the slightly different
  cosmologies.

\item {\bf merger trees: } The {\mun} model uses merger trees
  extracted from a dissipationless N-Body simulation
  \citep{Springel05}, {\mor} uses the Lagrangian semi-analytic code
  {\sc pinocchio} \citep{Monaco02}, and {\som} use a method based on
  the Extended Press-Schechter formalism, described in
  \citet{SomervilleKolatt99}.

\item {\bf substructure:} The {\mun} model explicitly follows dark
  matter substructures in the N-body simulation until tidal truncation
  and stripping reduce their mass below the resolution limit of the
  simulation \citep{DeLucia04a,Gao04}.  Beyond that point the merger
  time of the satellite is computed using the classical Chandrasekhar
  dynamical friction approximation (for more details see
  \citealt{DeLucia07b} and \citealt{DeLuciaHelmi08}). {\mor} and
  {\som} do not track explicitly dark matter substructures and assume
  satellite galaxies merge onto central galaxies after a dynamical
  friction time-scale which is assigned at the time the satellite
  enters the virial radius of the remnant structure, following
  \citet{Taffoni03} in the case of {\mor} and \citet{BoylanKolchin08}
  in the case of {\som}. The same models also account for tidal
  destruction of satellites.

\item {\bf cooling model: } {\mun} and {\som} use variations of the
  original cooling recipe of \citet{White91}, while {\mor} uses a
  modified model, described and tested against simulations in
  \citet{Viola08}, that predicts an enhanced cooling rate at the onset
  of cooling flows.

\item {\bf galaxy sizes, star formation, and SN feedback: } The three
  SAMs also differ in the details of the modeling of star formation,
  stellar feedback and galactic winds, as well as in the computation
  of galaxy sizes. We prefer not to discuss these processes in detail
  here and we refer the reader to the original papers for more
  details.

\item {\bf BH growth and AGN feedback: } In the {\mun} and {\som}
  models, the radio mode is fueled by accretion from the hot gas halo,
  and only haloes that can support a quasi-hydrostatic halo are
  subject to the radio mode heating (though the conditions used
  differ, see \citet{Croton06} and \citet{Somerville08}). In {\mor}
  the radio mode accretion comes from the cold gas reservoir
  surrounding the black hole. In the {\mun} and {\som} models, the
  ``bright mode'' or quasar mode is explicitly triggered by
  galaxy-galaxy mergers (though again, the details of the
  implementation differ), while in {\mor} it is associated with
  Eddington-limited high accretion rates (again coming from the cold
  reservoir). As noted above, {\mor} and {\som} include galactic scale
  AGN-driven winds associated with the bright mode, while {\mun} do
  not.

\end{itemize}

The three models were each normalized to fit a subset of low-redshift
observations.  The specific observations used, and the weight given to
different observations in choosing a favored normalization are
different for each of the three models, and we refer to the original
papers for details. The most important free parameters in all three
models are those controlling the efficiency of supernova feedback,
star formation, and ``radio mode'' AGN feedback. The efficiency of
supernova feedback is primarily constrained by the observed low-mass
slope of the stellar mass function ({\mor} and {\som}) or the
faint-end slope of the luminosity function ({\mun}). The efficiency of
star formation is mainly constrained by observations of gas fractions
in nearby spiral galaxies ({\mun} and {\som}) or by the cosmic SFR
density ({\mor}). The efficiency of the ``radio mode'' AGN feedback is
constrained by the bright or high-mass end of the observed LF or
SMF. Other important parameters are the effective yield of heavy
elements, which is constrained by the observed mass-metallicity
relation at $z=0$, and the efficiency of black hole growth, which is
constrained by the observed $z=0$ black hole mass vs. bulge mass
relationship.  We emphasize that we made no attempt to tune the models
to match each other, or to match any of the high redshift data that we
now compare with.

\section{Downsizing in stellar mass}
\label{sec:mass_down}

\begin{figure*}
  \centerline{
    \includegraphics[width=9cm]{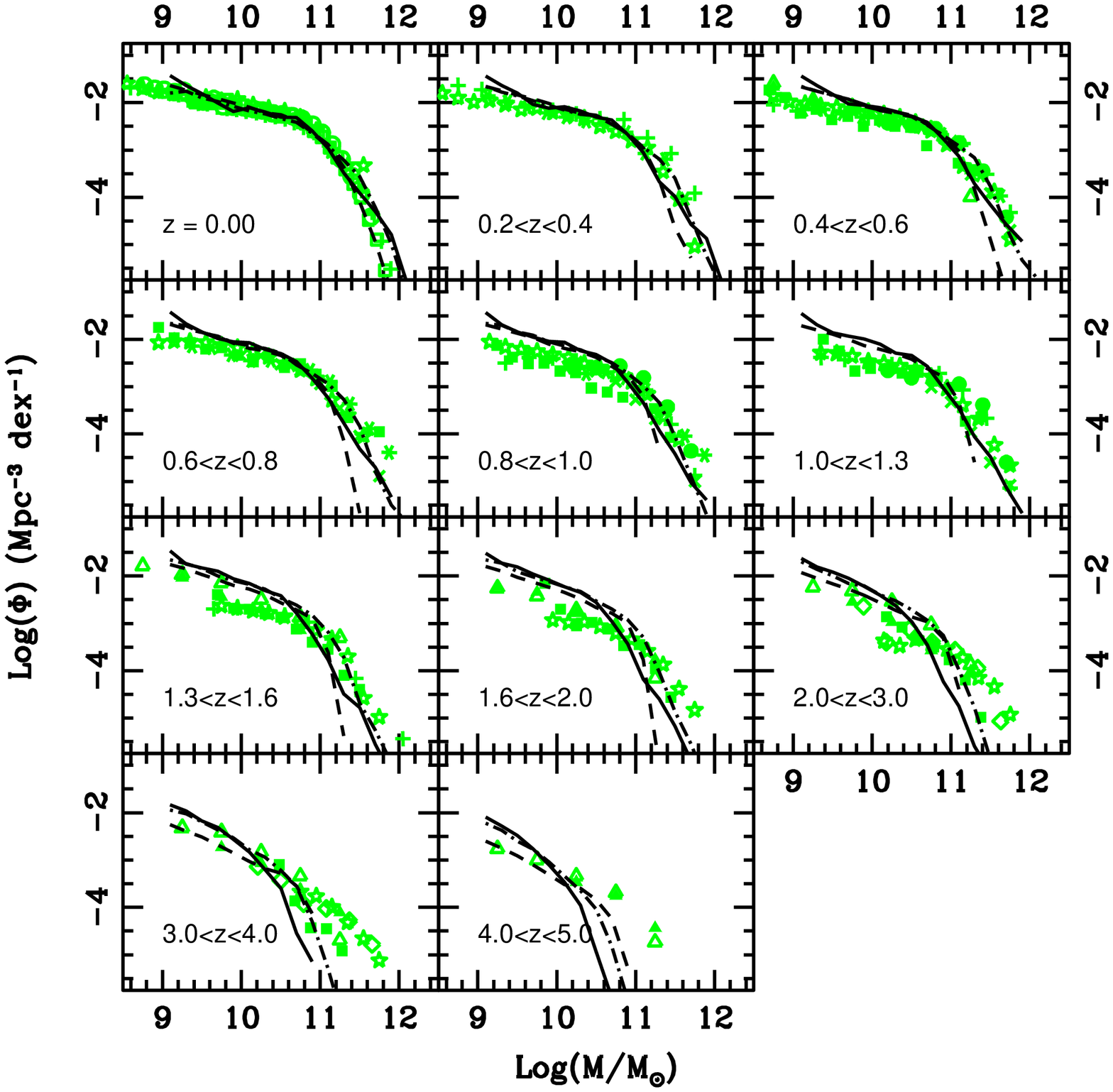} 
    \includegraphics[width=9cm]{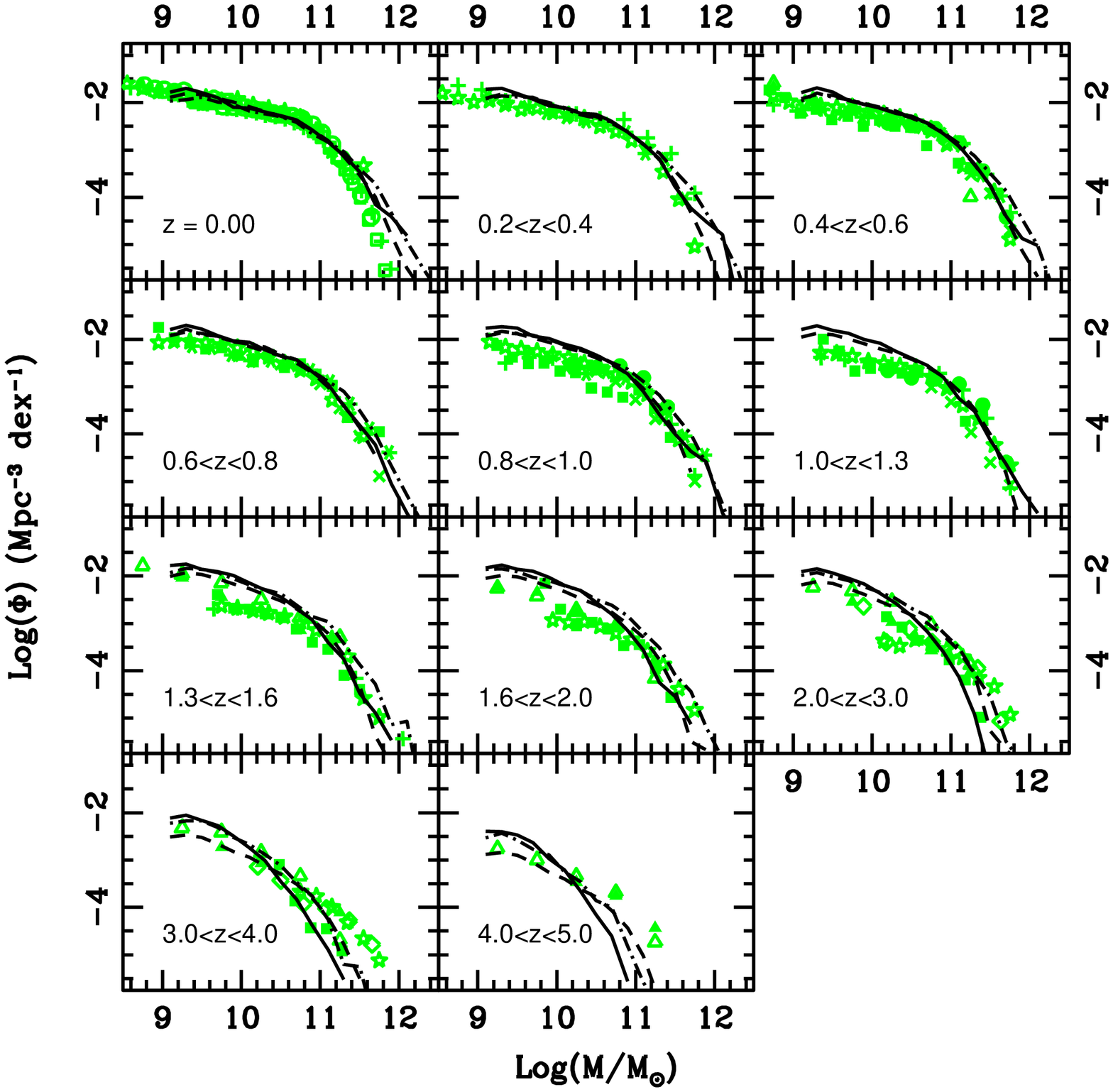} }
  \caption{Galaxy stellar MF as a function of redshift. Solid, dashed,
    and dot-dashed lines refer to the {\mor}, {\mun}, {\som} models
    respectively. In the right panels, all model predictions have been
    convolved with a Gaussian error distribution on $\log M_\star$
    with standard deviation of 0.25 dex. Green symbols correspond to
    observational measurements from \citet[SDSS, $z=0$]{Panter07},
    \citet[2MASS, $z=0$]{Cole01}, \citet[2MASS+SDSS, $z=0$]{Bell03},
    \citet[COMBO17, $0.2<z<1$]{Borch06}, \citet[Spitzer,
    $0<z<4$]{PerezGonzalez08}, \citet[DEEP2, $0.4<z<1.4$]{Bundy06},
    \citet[MUNICS, $0.4<z<1.2$]{Drory04}, \citet[FDF+GOODS,
    $0<z<5$]{Drory05}, \citet[GOODS-MUSIC, $0.4<z<4$]{Fontana06},
    \citet[VVDS, $0.05<z<2.5$]{Pozzetti07},
    \citet[MUSYC$+$FIRES$+$GOODS-CDFS, $1.3<z<4$]{Marchesini08}. All
    observational measurements have been converted to a Chabrier IMF,
    when necessary.}
  \label{fig:mf_evo}
\end{figure*}

In this section, we focus on the evolution of the galaxy stellar MF.
Our model predictions are compared with a compilation of published
observational estimates using different data-sets and methods to
compute stellar masses.  In the past, the rest-frame near infrared
light has been widely used as a tracer of the galaxy stellar mass
\citep{Cole01, Bell03}.  In more recent times, most mass estimates
\citep{Drory04, Drory05, Bundy06, Borch06, Fontana06, PerezGonzalez08,
  Marchesini08} have been based on multi-wavelength Spectral Energy
Distribution (SED) fitting algorithms. In this approach, broad-band
photometry is compared to a library of synthetic SEDs, covering a
relatively wide range of possible star formation histories,
metallicities, and dust attenuation values. A suitable algorithm is
then used to select the `best-fit' solution, thus simultaneously
determining photometric redshift, galaxy stellar mass and SFR. Stellar
mass estimates are therefore subject to several degeneracies (age,
metallicity, and dust) and their accuracy depends sensitively on the
library of star formation histories employed, and on the wavelength
range covered by observations \citep[see e.g.][]{Fontana04,
  Pozzetti07, Marchesini08}. In particular, most of these algorithms
assume relatively simple analytic star formation histories (with, in
some cases, some bursty star formation episodes superimposed), while
SAMs typically predict much more complex star formation histories,
with a non-monotonic behavior and erratic bursts. This may result in
certain biases in the physical parameters obtained from this method
\citep{Lee09}. Similarly, using different libraries of star formation
histories has an effect on the final mass determination
\citep{Pozzetti07,Stringer09}. Additional sources of uncertainty may
come from the physical ingredients in the adopted stellar population
models, e.g. to the treatment of particular stages of stellar
evolution, such as TP-AGB stars \citep{Maraston06,Tonini08}. Moreover,
due to the relatively small volumes probed at high redshift, cosmic
variance due to large-scale clustering is a significant source of
uncertainty, particularly in the number density of high-mass objects.

When high signal-to-noise spectroscopy is available, galaxy stellar
masses can be estimated by comparison of the observed spectra with
theoretical SEDs \citep[e.g.][for SDSS data]{Panter07}. In this case
the finer details of the spectrum can be used to give tighter
constraints on, e.g., stellar ages and metallicities. However, the
method is not free from uncertainties due to model degeneracies, and
contamination from AGN and/or strong emission lines (usually not
included in the theoretical SEDs) can in principle introduce
systematic errors or simply limit the accuracy of the mass estimate.

In Fig.~\ref{fig:mf_evo}, we show a compilation of different
observational measurements of the galaxy stellar MF from 2MASS
\citep{Cole01}, 2MASS$+$SDSS \citep{Bell03}, MUNICS \citep{Drory04},
FDF$+$GOODS \citep{Drory05}, COMBO17 \citep{Borch06}, DEEP2
\citep{Bundy06}, GOODS-MUSIC \citep{Fontana06}, SDSS \citep{Panter07},
VVDS \citep{Pozzetti07} Spitzer \citep{PerezGonzalez08},
MUSYC$+$FIRES$+$GOODS-CDFS \citep{Marchesini08} (green points; left
and right panels show the same data).  All estimates have been
converted to a common (Chabrier) IMF when necessary; we use a factor
of 0.25 dex to convert from Salpeter to Chabrier. These stellar mass
functions are fairly consistent among themselves, but the scatter
becomes larger at higher redshift, in particular for the high-mass
tail (which is significantly affected by cosmic variance).

A note on the errors and uncertainties associated with these
observationally derived stellar mass functions is in order. Most
published papers quote only Poisson errors on their mass function
estimates. However, as noted above, both systematic and random errors
can arise from the unknown true star formation histories,
metallicities, and dust corrections, and also from photometric
redshift errors, differences in stellar population models, the unknown
stellar IMF and its evolution, and cosmic
variance. \citet{Marchesini08} carry out an extensive investigation of
the impact of all of these sources of uncertainty on their derived
stellar mass functions. In their Figures~13 and 14 they show a
comparison of their results, including these comprehensive error
estimates, with the three models presented here.  Their analysis shows
that the evidence for differential evolution in the stellar mass
function at $z<2$, with more massive galaxies evolving more slowly
than less massive ones, becomes weak when all sources of uncertainty
in the stellar mass estimates are considered. When this is done, the
observed evolution appears to be consistent with pure density
evolution.

\begin{figure}
  \centerline{
    \includegraphics[width=9cm]{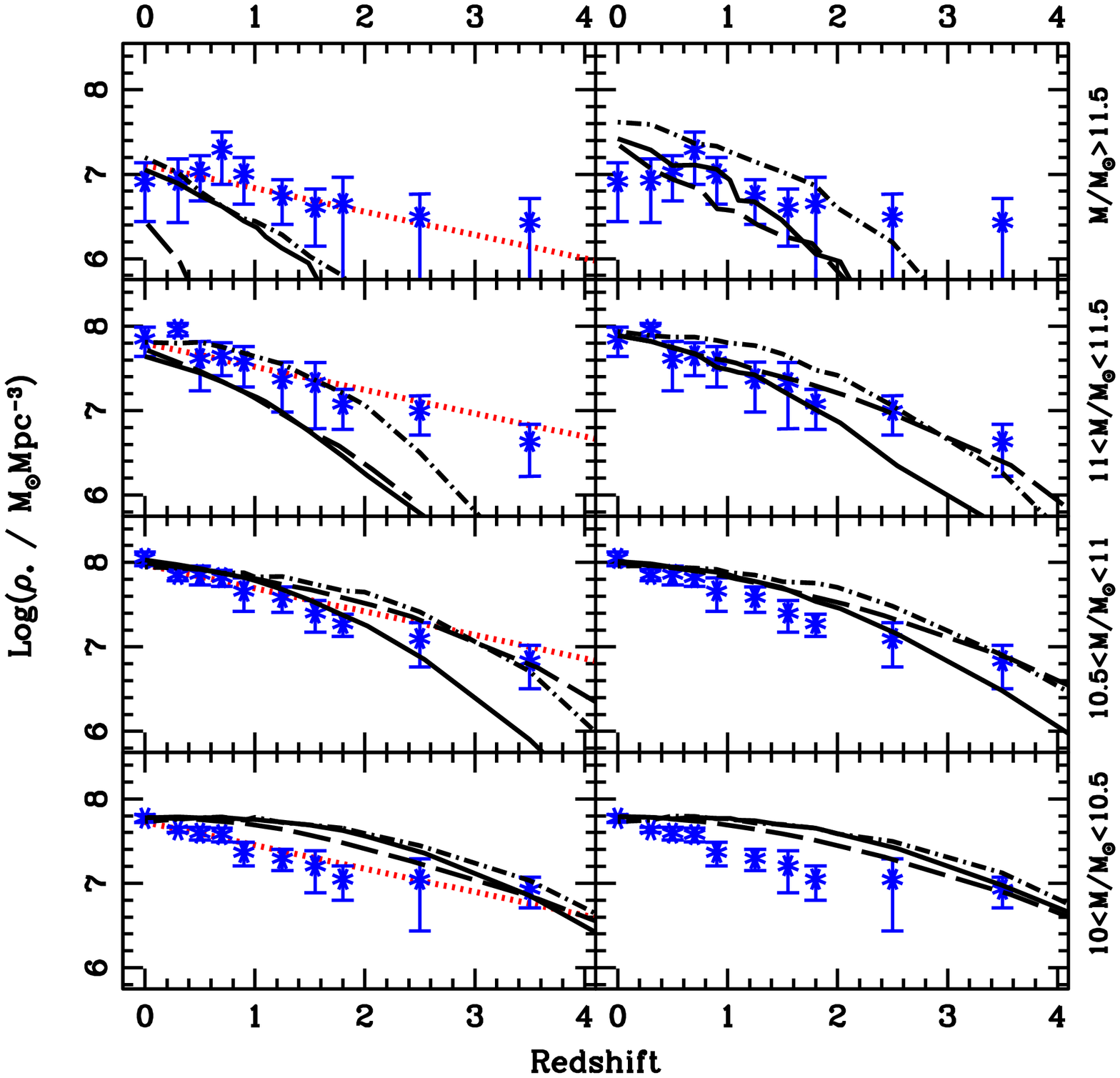} }
  \caption{Stellar mass density in bins of galaxy stellar mass, as a
    function of redshift. Blue asterisks are estimated from available
    observational estimates of the galaxy stellar MF (see text for
    details). The thick red line in the left panels shows the expected
    evolution in a pure density evolution scenario. Solid, dashed, and
    dot-dashed lines refer to the {\mor}, {\mun}, and {\som} models
    respectively.  In the right panels, the model predictions have
    been convolved with a Gaussian error on $\log M_\star$ with a
    standard deviation of 0.25 dex.}
  \label{fig:smd_evo}
\end{figure}

In order to make DS in stellar mass more evident, we divide galaxies in bins of
stellar mass and, by averaging over the MF estimates of Fig~\ref{fig:mf_evo},
compute the stellar mass density of galaxies in these stellar mass bins as a
function of redshift.  These stellar mass densities agree with estimates
published by \citet{Conselice07} and \citet{Cowie08} and are shown in
Fig.~\ref{fig:smd_evo} (left and right panels contain the same data). The
quoted errors refer only to the scatter between the estimates from different
samples (note that this scatter is larger than the quoted errors on individual
determinations, confirming that these errors are underestimated, as discussed
above).

DS in stellar mass should consist of a differential growth of stellar
mass density, such that massive galaxies are assembled earlier and
more rapidly than low mass galaxies. Examining Fig.~\ref{fig:smd_evo},
it is hard to claim convincing evidence for such a behavior from
these data: although the evidence for the growth of stellar mass
density is clear, its rate of growth is very similar for all mass
bins.  To further illustrate this point we perform a linear regression
of the ${\rm Log} \rho_\star - z$ relation in each mass bin; the
slopes we obtain are consistent within their statistical errors. We
then rescale the densities to the $z=0$ value of their regressions and
fit the whole sample, obtaining the thick dotted line in the left
panels of Fig.~\ref{fig:smd_evo}. The fit, valid in the $10^9<M_\star
/ \msun <10^{12}$ range, has a $\chi^2$ probability of $>95\%$).

The predicted stellar MFs for the {\mor}, {\mun} and {\som} models
(solid, dashed, and dot-dashed lines respectively) are shown in the
left panels of Fig.~\ref{fig:mf_evo} and \ref{fig:smd_evo}, while in
the right panels of the two figures, the model stellar masses have
been convolved with a statistical error on $\log M_\star$.  As we have
discussed above, this error distribution depends on many factors, such
as the specific algorithm and stellar population models used to
estimate stellar masses for each sample, the magnitude and redshift of
the galaxy, and characteristics of each observational survey, such as
the volume covered and the number and wavelength coverage of the
photometric bands that are available. A detailed accounting of this
complex error distribution for each observational dataset is clearly
beyond the scope of this paper. Rather than simply ignoring the impact
of errors in the stellar mass estimates on our data-model comparison,
as has usually been done in the past, we adopt a simple approach that
is meant to be illustrative rather than definitive. We assume that the
error has a Gaussian distribution {\it (independent of mass and
  redshift)} with a standard deviation of $0.25$ dex.  This assumed
uncertainty roughly corresponds to the mean value of the formal error
in the stellar mass determination from the GOODS-MUSIC catalogue
\citep[][their Figure~2]{Fontana06}, is lower (by about 0.1 dex) than
that estimated by \citet{Bundy06}, and is roughly consistent with the
findings of \citet{Stringer09}.

The first thing to note is that the models give fairly consistent
predictions. Secondly, as redshift increases, the intrinsic model
predictions (i.e without convolution with errors) show a significant
deficit of massive galaxies (the two bins $10^{11}<M_\star/\msun
<10^{11.5}$ and $M_{\star}>10^{11.5}\ \msun$) with respect to the
data.  The error convolution does not affect the power-law part of the
MF, but it has a significant impact upon its high mass tail, as
already pointed out by \citet{Baugh06}, \citet{Kitzbichler07}, and
more recently by \citet{Stringer09}. Because the models were tuned to
match the $z=0$ stellar MF or LF {\em without} errors, this
convolution causes a small apparent overestimate of the number of the
most massive galaxies at $z\sim0$. This could be corrected by tuning
the radio mode AGN feedback in the models. However, there are
indications that the observed magnitudes and stellar masses of the
brightest local galaxies may be underestimated by significant amounts
(see the discussion in S08). Therefore we do not re-tune the models to
correct this apparent discrepancy. When these observational
uncertainties are taken into account, model predictions for massive
galaxies are in fairly good agreement with observations over the
entire redshift interval probed by the surveys that we considered
(with {\mor} being $\sim2\sigma$ low at $z>2$).

In lower stellar mass bins ($10^{10.5}< M_\star/\msun < 10^{11}$ and
in particular $10^{10} < M_\star/\msun < 10^{10.5}$), all three models
overpredict the observed stellar MF and stellar mass density at high
redshift ($z\gtrsim 0.5$), with the discrepancy increasing with
increasing redshift. Thus, a robust prediction of the models seems to
be that the evolution of less massive galaxies is slower than that of
more massive ones --- i.e., the models do not predict stellar mass
``downsizing'' but rather the opposite behavior (sometimes called
``upsizing''). This discrepancy has already been noted in previous
papers \citep{Fontana04,Fontanot07b}, and extends to other
models. Therefore, the models seem to be discrepant with observations
even if the real universe shows mass-independent density evolution
rather than downsizing. As a caveat, it is worth mentioning that,
although the different groups have performed detailed completeness
corrections, it is still possible that the high redshift samples may
be incomplete at the lowest stellar masses. In Fig.~\ref{fig:mf_evo}
and~\ref{fig:smd_evo} we show data only for mass ranges where the
corresponding authors claim that no completeness correction is
necessary.

\begin{figure*}
\centerline{
    \includegraphics[width=9cm]{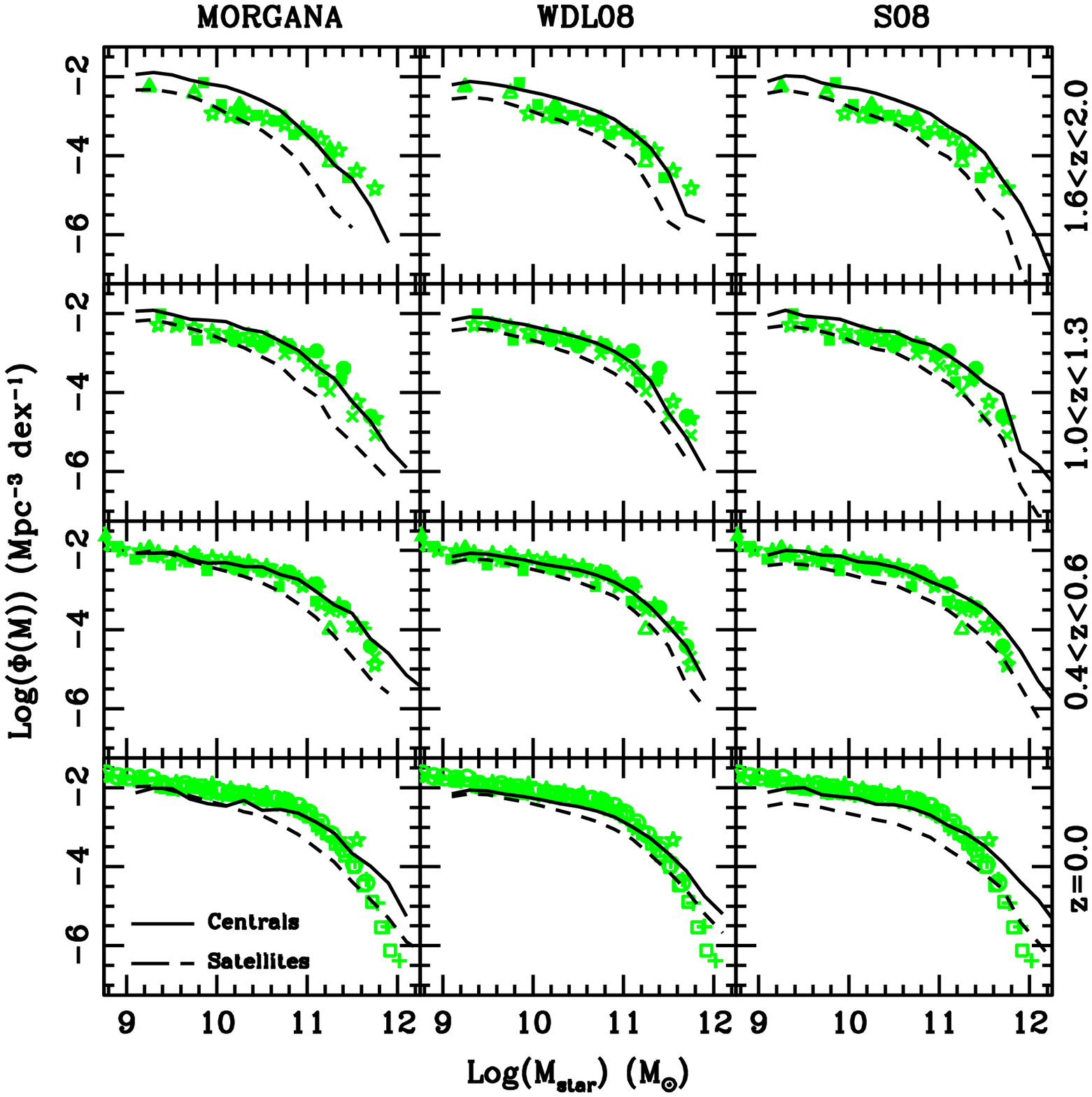} 
    \includegraphics[width=9cm]{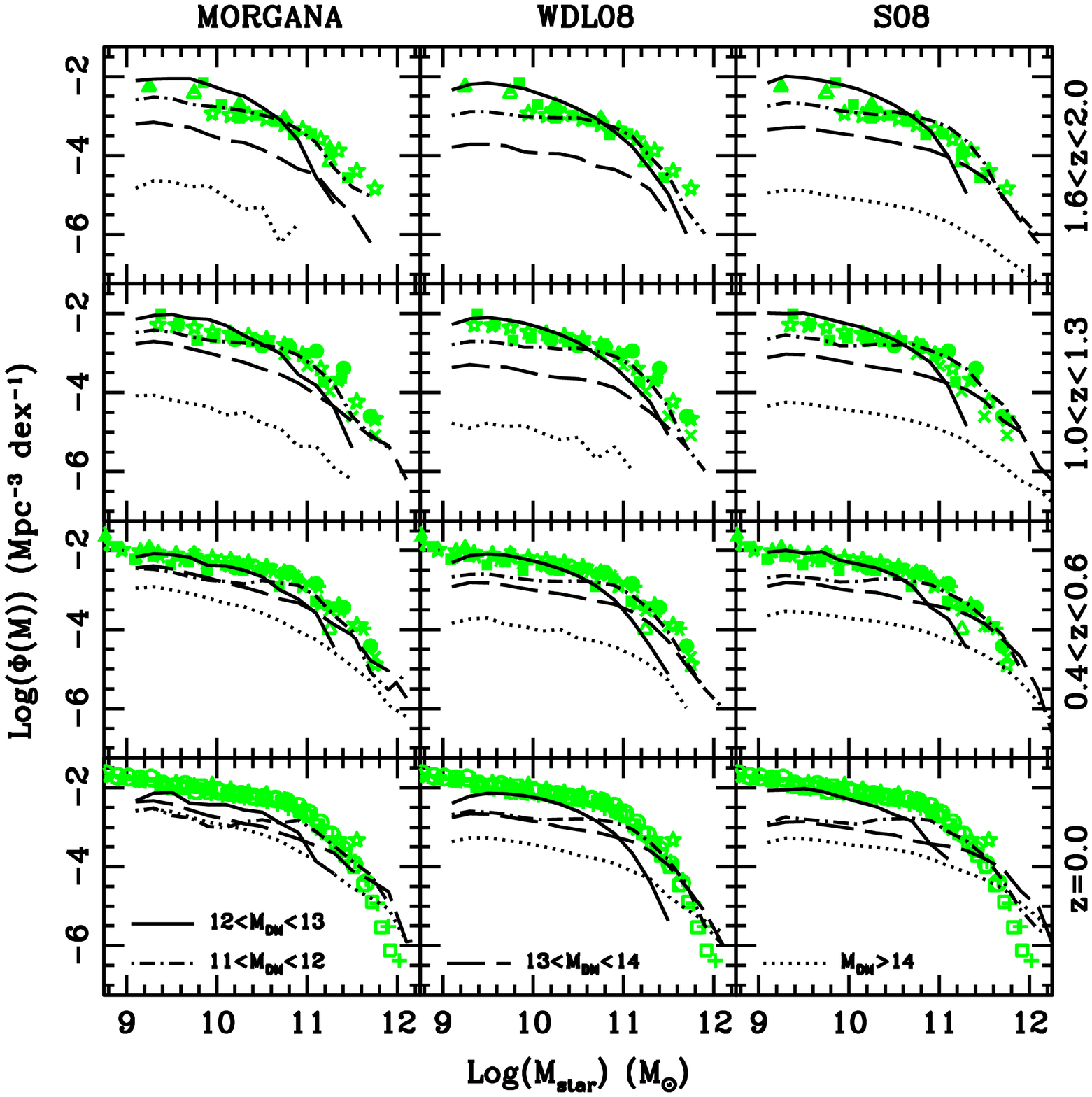} }
  \caption{Left panel: solid and dashed lines refer to model galaxy
    SMFs for central and satellite galaxies respectively. Right panel:
    model galaxy MFs in bins of parent DM halo mass. Different models
    are shown in different columns, as labeled. In both panels,
    observational measurements of the global SMFs at different
    redshifts are shown as in fig.~\ref{fig:mf_evo}.}
  \label{fig:cen_sat}
\end{figure*}

In order to further investigate the evolution of the predicted galaxy
SMF, we consider separately the contribution from central and
satellite galaxies at different redshifts. Model predictions
(convolved with observational uncertainties as before) are shown in
the left panels of Fig.~\ref{fig:cen_sat}. As a reference, in each
panel we show the {\it total observed} MFs at the considered
redshift. The three models predict a similar evolution for the two
subpopulations. It is evident that central galaxies are the main
contributors to the overprediction of low-mass galaxies at $z<2$: the
models predicts a roughly constant $z<2$ number density of low-mass
central galaxies, while the low-mass satellite population shows a
gradual increase which is due to the infall of field galaxies into
galaxy groups and clusters.  This implies that small objects are
over-produced while they are central galaxies, and the excess is not
primarily due to inaccuracies in the modeling of satellites.

In the right panels of Fig.~\ref{fig:cen_sat} we show the evolution of
the galaxy SMF split in bins of parent halo mass at the considered
redshift. Again, the three SAMs predict similar trends: it is evident
that the main contributors at all redshifts to the low-mass end excess
are low-mass ($10^{11}<M_h/\msun<10^{12}$) dark matter halos. These
results suggest that, in order to cure the discrepancies seen in these
three models and others, we should seek a physical process that can
suppress star formation in central galaxies hosted by intermediate to
low-mass haloes ($M_h/\msun<10^{12}$).

\section{Downsizing in Star Formation Rate}
\label{sec:sfr_down}

\begin{figure*}
\centerline{
    \includegraphics[width=17cm]{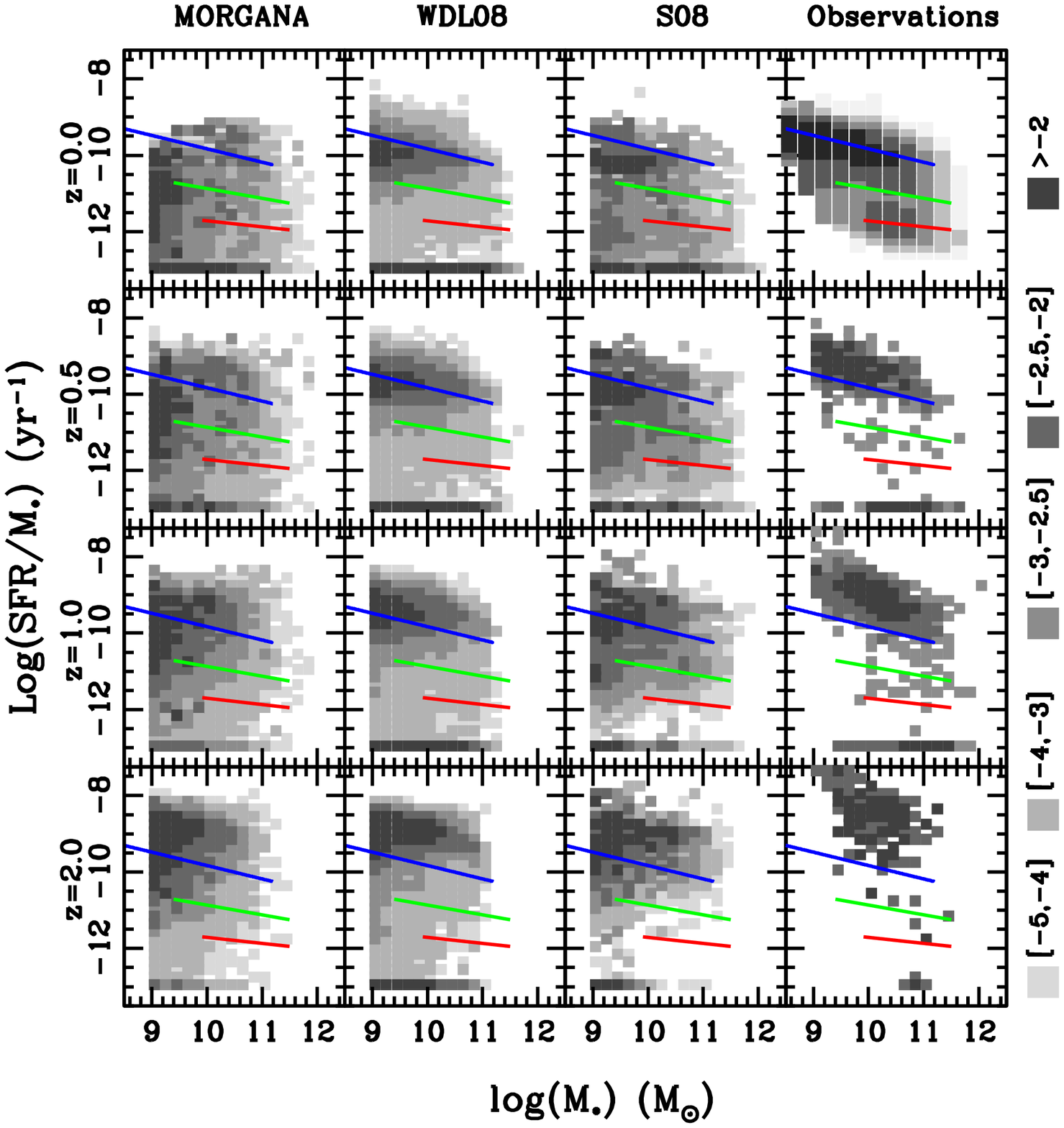} 
    }
    \caption{The 2-dimensional distribution of specific star formation
      rates versus stellar mass for the three different SAMs at
      different redshifts, and for a compilation of observations.
      Model galaxies with SSFR $ < 10^{-13}\ {\rm yr}^{-1}$ have been
      assigned SSFR $ = 10^{-13}\ {\rm yr}^{-1}$. Observational data
      are from \citet{Schiminovich07}, \citet{Noeske07} and
      \citet{Santini09}. Blue, red and green lines indicate the
      observed $z=0$ ``star-forming'' sequence, the ``quenched''
      sequence and the ``green valley'', respectively.}
  \label{fig:2dd}
\end{figure*}

\begin{figure*}
\centerline{
    \includegraphics[width=17cm]{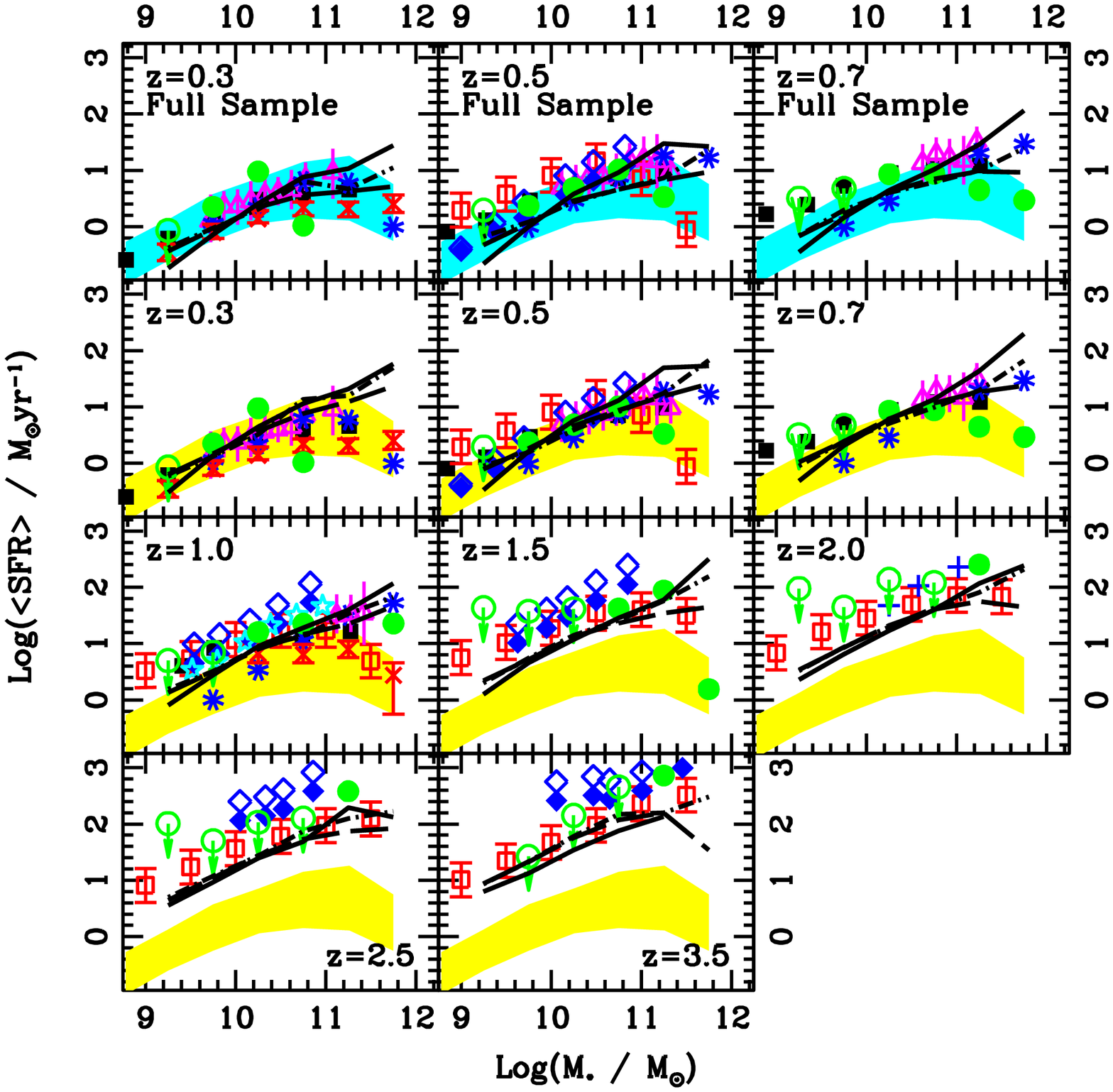} 
  }
  \caption{Average SFR of galaxies in bins of stellar mass and
    redshift.  Data are from \citet[red open squares]{Drory08},
    \citet[black filled squares]{Bell07}, \citet[magenta open
    triangles]{Noeske07}, \citet[red crosses]{Chen09}, \citet[blue
    asterisks]{Martin07}, \citet[cyan stars]{Elbaz07}, \citet[blue
    crosses]{Daddi07}, \citet[filled and open blue diamonds]{Dunne08},
    and \citet[green filled circles, open circles indicate upper
    limits]{Santini09}.  Errorbars are shown where provided by the
    authors. The shaded area represents the confidence region for the
    lowest redshift bin.  Solid, dashed and dot-dashed lines refer to
    the {\mor}, {\mun} and {\som} models respectively.  Model
    predictions have been convolved with the errors on stellar mass
    and SFR. Top row: all model galaxies are included in the average;
    second through fourth rows: Only active (SSFR $> 10^{-11}\ {\rm
      yr}^{-1}$) model galaxies have been included in the average. The
    results for active and all galaxies are nearly indistinguishable
    for the high redshift ($z \geq 1$) bins, which is why we do not
    show both cases. }
  \label{fig:msf_evo}
\end{figure*}

The downsizing in galaxy SFR seen in lookback studies most closely
corresponds to the original definition of DS. There are, however,
several different forms in which the diagnostics of DS in SFR may been
cast observationally. In addition, one can identify two different
trends that might be called ``downsizing'': first, the normalization
of the ``star forming sequence'' of galaxies shifts downwards with
decreasing redshift; second, galaxies move off the star forming
sequence and become `quenched' or passive as time progresses. These
different behaviors may offer clues to the physical mechanisms
responsible, e.g., the downward shift of the SF sequence might be due
to simple gas exhaustion by star formation, while ``quenching'' is
presumably due to a more dramatic process such as AGN feedback. If
downsizing is occuring, this evolution should happen in a differential
way, with more massive objects being quenched earlier and/or more
rapidly. In order to probe these different possible ``paths'' for
downsizing, we will consider several different ways of slicing and
plotting the distribution function of SFR as a function of stellar
mass and redshift: 1) the 2-d distribution of stellar mass and SFR in
several redshift bins (2) the average SFR as a function of stellar
mass, plotted in redshift bins (3) the SFR density contributed by
objects of different stellar masses, as a function of redshift (4) the
evolution of the stellar mass function of active vs. passive
galaxies. 

Star formation rates are estimated using different observational
tracers, such as H-$\alpha$ emission lines, UV, mid- and far-IR
emission, and radio. SFR may also be estimated by fitting SF histories
to multi-wavelength broadband SED's, in a similar manner used to
estimated stellar masses. SFR estimates are impacted by many of the
same sources of uncertainty as stellar mass estimates (such as
propagated errors from photometric redshift uncertainties and
sensitivity to the assumed stellar population models, stellar initial
mass function, and star formation histories), and also each tracer
carries its own set of potential problems. For example, SFR estimates
based on emission lines such as H-$\alpha$ are metallicity dependent
and (typically fairly large) corrections for dust extinction must be
applied. A potential advantage to this approach is that dust
corrections can be fairly reliably estimated from the Balmer
decrement; however, these measurements are currently impractical at
high redshift as they would require highly multiplexed, deep NIR
spectroscopy. SFR estimates based on the UV continuum alone suffer
mainly from the very large and uncertain dust corrections (extinction
estimates based on the UV spectral slope, while widely used, are quite
uncertain). Estimates based on the mid-IR (e.g. 24$\mu$) suffer from
highly uncertain k-corrections (as strong PAH features move through
the observed bandpass), potential strong contamination by AGN,
uncertainties in the IR SED templates (due to our lack of knowledge
about the composition and state of the emitting dust), and possibly
contamination by heating from old stellar populations. Measurements of
the longer wavelength thermal IR, near the peak of the dust emission
($\sim 100 \mu$) offer perhaps the most promising approach to
obtaining robust estimates of total SFR. These are, however, currently
available only for a small number of very IR-luminous galaxies.
Moreover, all these indicators are usually calibrated on local galaxy
samples, and the systematics connected with applying them to higher
redshift are poorly known.

The observed SFRs used in this Section have been obtained from UV +
Spitzer 24$\mu$m \citep{Bell07,Zheng07}, Spitzer 24$\mu$m
\citep{Conselice07}, GALEX FUV \citep{Schiminovich07}, emission lines
+ Spitzer 24$\mu$m \citep{Noeske07}, SED-fitting continuum at 2800
{\AA} \citep{Drory08,Mobasher09}, GALEX FUV + Spitzer 24$\mu$m
\citep{Martin07}, Balmer absorption lines \citep{Chen09}, SED-fitting
+ Spitzer 24$\mu$m \citep{Santini09}, and radio \citep{Dunne08}.

The comparison of models and data is also made difficult by the
complex selection criteria involved. Most SFR estimates used here have
poor sensitivity to sources with low SFRs, leading to many upper
limits; for instance, SFR estimates for passive objects are poorly
constrained by SED fitting techniques.  Several authors have attempted
to correct for incompleteness by stacking images of objects with
similar masses to obtain deeper detections, or by using only galaxies
with active star formation to compute the average.  A proper
comparison should take into account the selection effects of each
dataset; however, systematics are large and poorly understood, so a
detailed comparison at this stage is of doubtful utility.  With all
these caveats in mind, we compare our models to the data {\em at face
  value}, trying again to assess whether DS is seen in the data and to
what extent models are consistent with available observations.
Moreover, analogously with stellar masses, we convolve model SFRs with
a log-normal error distribution; for its amplitude we use a value of
0.3 dex, roughly equal to the median formal error of SFRs in
GOODS-MUSIC.  In the light of what said above, this estimate is
clearly naive, but it allows us to determine the gross effect of
(random) uncertainties in SFR determinations.  We find that our
results are fairly insensitive to the inclusion of this error.

In Fig.~\ref{fig:2dd} we show the 2-d distribution of specific SFR
(SSFR) as a function of stellar mass, for several redshifts from $z=2$
to 0, for all three models and for a compilation of observational data
\citep{Schiminovich07, Noeske07, Santini09}. All model galaxies with
SSFR$ < 10^{-13}\ {\rm yr}^{-1}$ have been assigned SSFR $ =
10^{-13}\ {\rm yr}^{-1}$ (this causes the thin quenched sequence at
the bottom of each panel). In each panel, we plot the locations of the
``star forming'' and ``quenched'' sequences at $z\sim 0$ from local
observations based on SDSS+GALEX \citep{Salim07,Schiminovich07}, and
of the so-called ``green valley'' that divides the two sequences. We see
that all three models show qualitatively similar behavior. Perhaps the
clearest discrepancy between the models and data is that the SSFR of
low-mass galaxies in all three models are the same as or, in the case
of {\mor}, even lower than those of massive galaxies, while in the
observations a clear trend is seen such that lower mass galaxies have
higher SSFR. In the models, the slope of the SF sequence does not
appear to change significantly over time between $z\sim2$ and $z\sim
0$, while the normalization of this sequence decreases over time. Also
in all three models, there are few if any massive passive galaxies in
place at $z\sim2$; it remains to be seen whether this is in conflict with
observations.

In Figure~\ref{fig:msf_evo} we show the evolution of the average SFR
of galaxies as a function of stellar mass, for eight redshift bins
from $z\sim0.3$ to $\sim3.5$; data are taken from \citet{Bell07},
\citet{Noeske07}, \citet{Martin07}, \citet{Drory08}, \citet{Chen09},
\citet{Santini09} and \citet{Dunne08}.  For the GOODS-MUSIC data
\citep{Santini09} we show with open (filled) circles the bins where,
according to the authors, the incompleteness is (is not) significant;
open symbols are then upper limits to the average SFR.  In order to
illustrate the redshift evolution of this quantity, in all panels the
shaded cyan/yellow area represents the confidence region of the
$z\sim0$ observations. In the top row, we show redshifts 0.3--0.7, and
show model results in which we average over all galaxies. In the
second row, we repeat the redshift bins 0.3, 0.5, and 0.7 but this
time include in the model averages only star-forming galaxies (defined
here as having SSFR$> 10^{-11}\ {\rm yr}^{-1}$). The remaining panels
show model averages for active galaxies only, for higher redshifts $1
< z < 3.5$. We only show the low redshift bins for both active and all
galaxies because it is only at these redshifts that there are any
significant difference in the results. We can see, however, that at
low redshift, the inclusion of passive galaxies causes a turn-over in
the average SFR at high masses in the {\mun} and {\som} models.

We first note again the good agreement between the results of the
three different SAMs seen in figure~\ref{fig:msf_evo}, a result that
we did not necessarily expect given the different implementations of
star formation and feedback. Regarding the comparison with
observations, we find that the average SFR of {\em low-mass} galaxies
($M_{\star} \lesssim 10^{11} \msun$) is underestimated by the models
at all redshifts, as we already noted from Figure~\ref{fig:2dd}. The
average star formation rates for massive galaxies generally lie near
the middle of the range of different observational estimates at low
redshift, and near the lower envelope of observational estimates at
higher redshift ($z \gtrsim 2$). Several previous studies
\citep{Elbaz07,Daddi07,Santini09} compared the predictions of a
slightly different version of the {\mun} models with a single
observational estimate of the SFR as a function of stellar
mass. \citet{Elbaz07} found that the model predictions were lower than
their observational estimates at $z\sim1$ by about a factor of two,
while \citet{Daddi07} and \citet{Santini09} found that the models were
low by a factor of $\sim 5$ at $z\sim2$. Our results are entirely
consistent with their findings, but we also see that (as already noted
above) the dispersion in different observational estimates of the
average SFR at fixed stellar mass is as large, or larger, than the
discrepancy between the model predictions and the observational
estimates of these previous studies.

{\mor} produces too few massive, passive galaxies at late times,
resulting in an overestimate of the SFR of massive objects at low
redshift. This was studied in more detail in \cite{Kimm08}, and is due
to a less efficient, or delayed, quenching of the cooling flows in
massive haloes via radio mode feedback.

\begin{figure}
\centerline{
    \includegraphics[width=9cm]{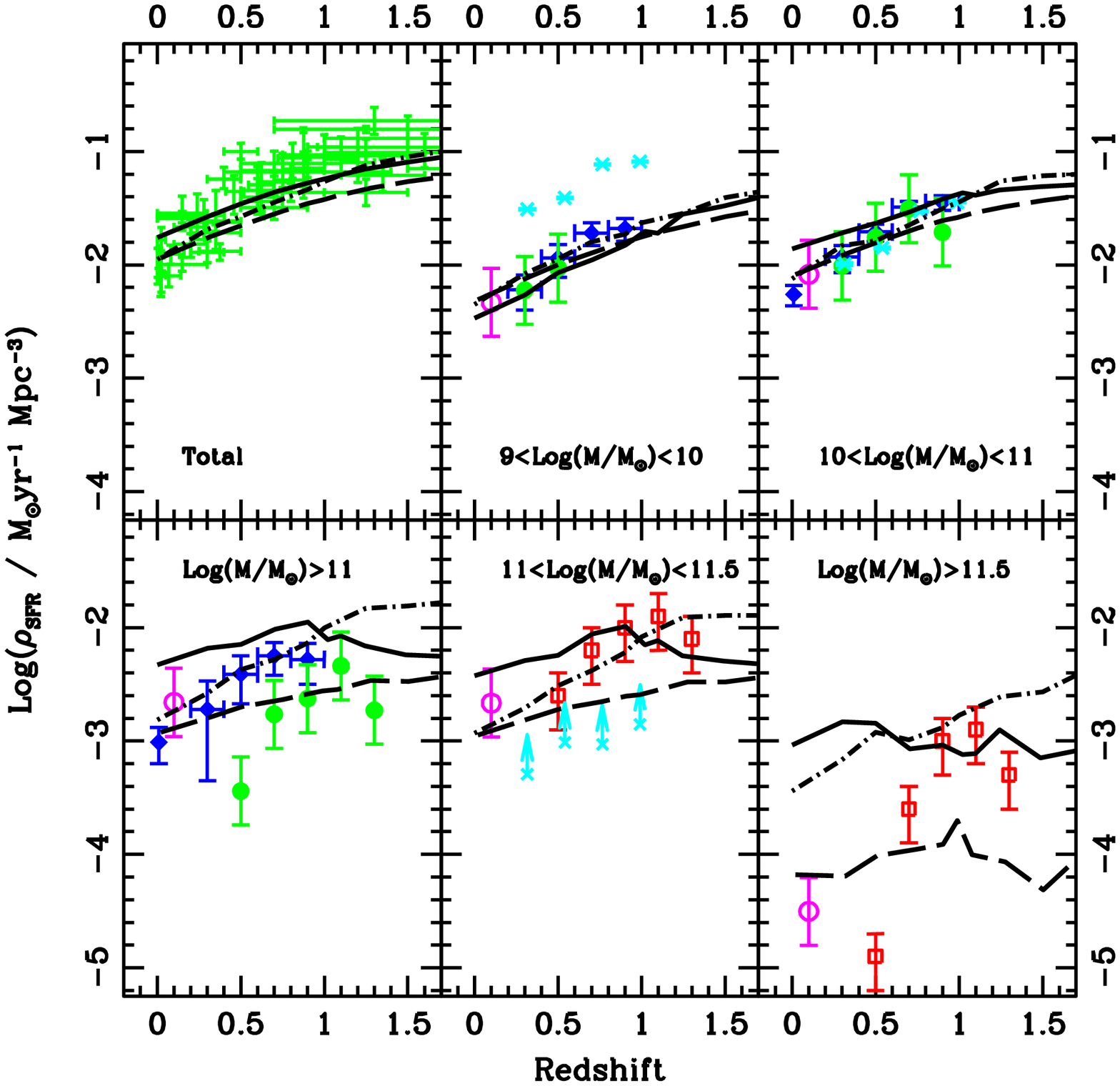} 
    }
    \caption{SFR density contributed by galaxies of different stellar mass
      (stellar mass is measured at the redshift of the plotted points). Data
      are from \citet[blue filled diamonds]{Zheng07}; \citet[cyan
        stars]{Mobasher09}; \citet[magenta open circles]{Schiminovich07};
      \citet[red open squares]{Conselice07} and from GOODS-MUSIC (green open
      and filled circles).  Solid, dashed and dot-dashed lines refer to the
      {\mor}, {\mun}, and {\som} models respectively.  Model predictions have
      been convolved with errors in stellar mass and SFR as explained in the
      text. Note that the lower panels do not represent a mass sequence, in
      order to compare model predictions with observational determinations.}
  \label{fig:csfr}
\end{figure}

Similar conclusions can be reached by considering the SFR density, as
a function of redshift, contributed by galaxies of different stellar
mass (Fig.~\ref{fig:csfr}). We used the K-band selected GOODS-MUSIC
catalogue, complete to $K<23.5$ \citep{Grazian06}, to compute the SFR
density as a function of stellar mass.  Following the discussion in
\citet{Fontana06}, we translated the magnitude limit into a stellar
mass limit, and computed SFRs either with SED fitting using photometry
from the NUV to the mid-IR, or with Spitzer 24$\mu$m fluxes when
available. We also plot several other data sets from the literature:
local points from SDSS+GALEX from \citet{Schiminovich07}; the results
of \citet{Conselice07}, based on the Palomar/DEEP2 Survey; estimates
from stacked 24 $\mu$m flux from the COMBO-17 survey \citep{Zheng07};
and estimates from UV luminosity alone \citep{Mobasher09}.

The three models again give fairly consistent results, although the
predictions diverge in the higher mass bins. All three models show a
gentle decline in the SFR density for low mass galaxies, and if
anything a somewhat flatter behavior for the SFR density in massive
galaxies. This time the model predictions agree well with the
observations for small mass galaxies, because the higher number of
small galaxies compensates for the lower SSFR of the objects.

\begin{figure}
\centerline{
    \includegraphics[width=9cm]{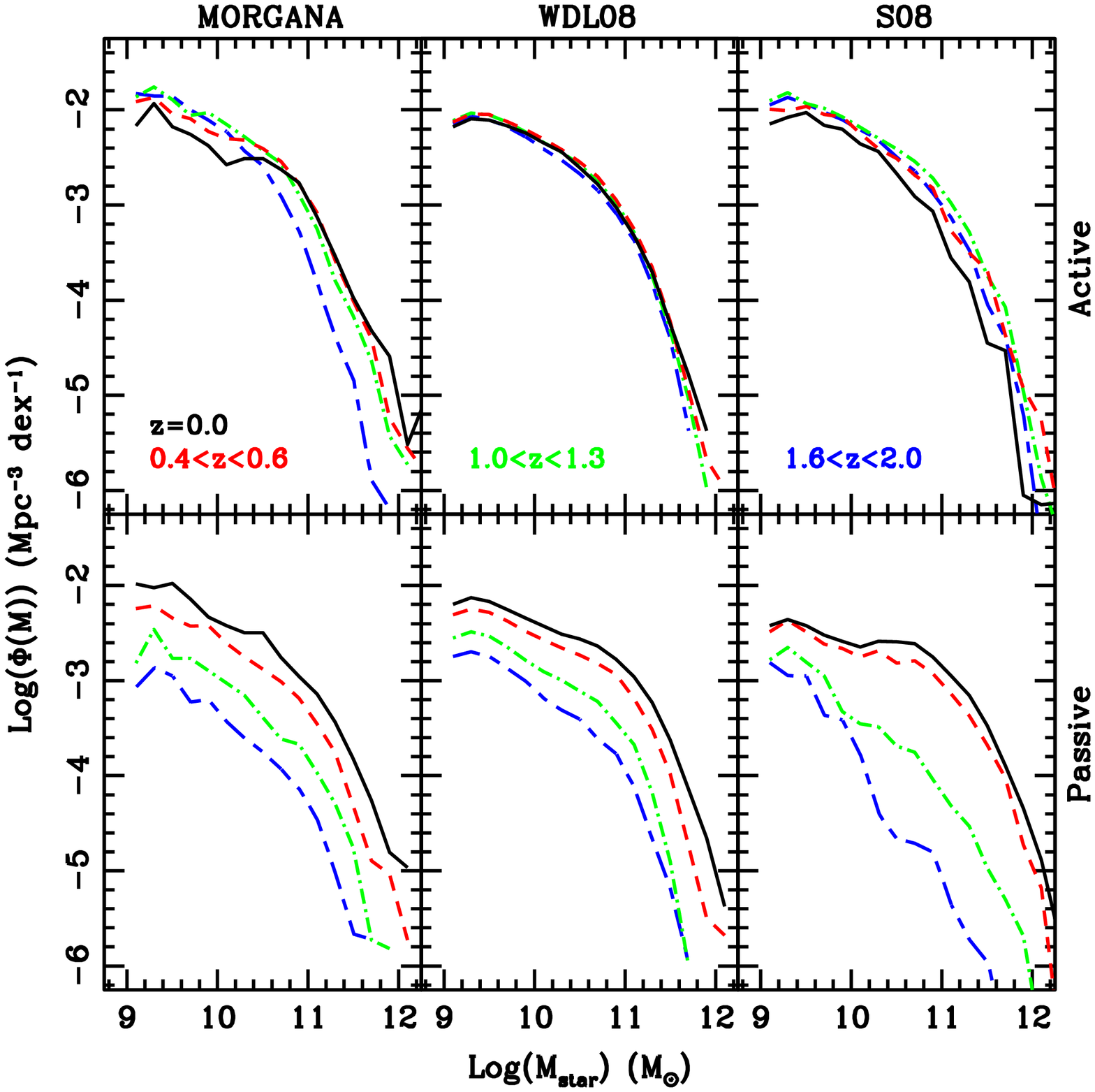} }
  \caption{Stellar MFs of active (SSFR$> 10^{-11}\ {\rm yr}^{-1}$, upper
    panels) and passive (SSFR$< 10^{-11}\ {\rm yr}^{-1}$, lower panels)
    galaxies.  Solid, dashed, dot-dashed and long-short dashed lines refer to
    $z=0$, $0.4<z<0.6$, $1.0<z<1.3$ and $1.6<z<2.0$, respectively. Different
    columns show the three different models, as labeled.}
  \label{fig:act_pas}
\end{figure}

Another way to characterize DS in star formation is by dividing
galaxies into active (blue) and passive (red) populations, then
computing the two stellar MFs or, alternatively, the $K$-band
luminosity functions.  As pointed out by \citet{Borch06}, using the
COMBO17 sample, and \citet{Bundy06}, using DEEP2, the two mass
functions cross at a characteristic mass which grows with redshift.
Instead of using the color criterion we divide our sample into passive
and active galaxies using a threshold value for the SSFR of $10^{-11}\
{\rm yr}^{-1}$ \citep{Brinchmann04}. Figure~\ref{fig:act_pas} shows
the evolution of the stellar MFs of active and passive galaxies as
predicted by the three models. The MF of active galaxies show almost
no evolution since $z\sim2$, whereas most of the evolution of the MF
is due to the build-up of the passive population; this is
qualitatively consistent with the observational results. However,
observations \citep{Bundy06, Borch06} show that the stellar MF of red
(passive) galaxies peaks at $\sim10^{11} \msun$ and decreases at lower
masses. In other words, in observed samples, low mass galaxies are
predominantly blue (active), while in our models the low-mass slope of
the SMF is nearly the same for active and passive galaxies. This
result still holds when galaxies are divided using colours rather than
SSFR, and this marks another discrepancy between models and data for
small galaxies.

\begin{figure}
\centerline{
  \includegraphics[width=9cm]{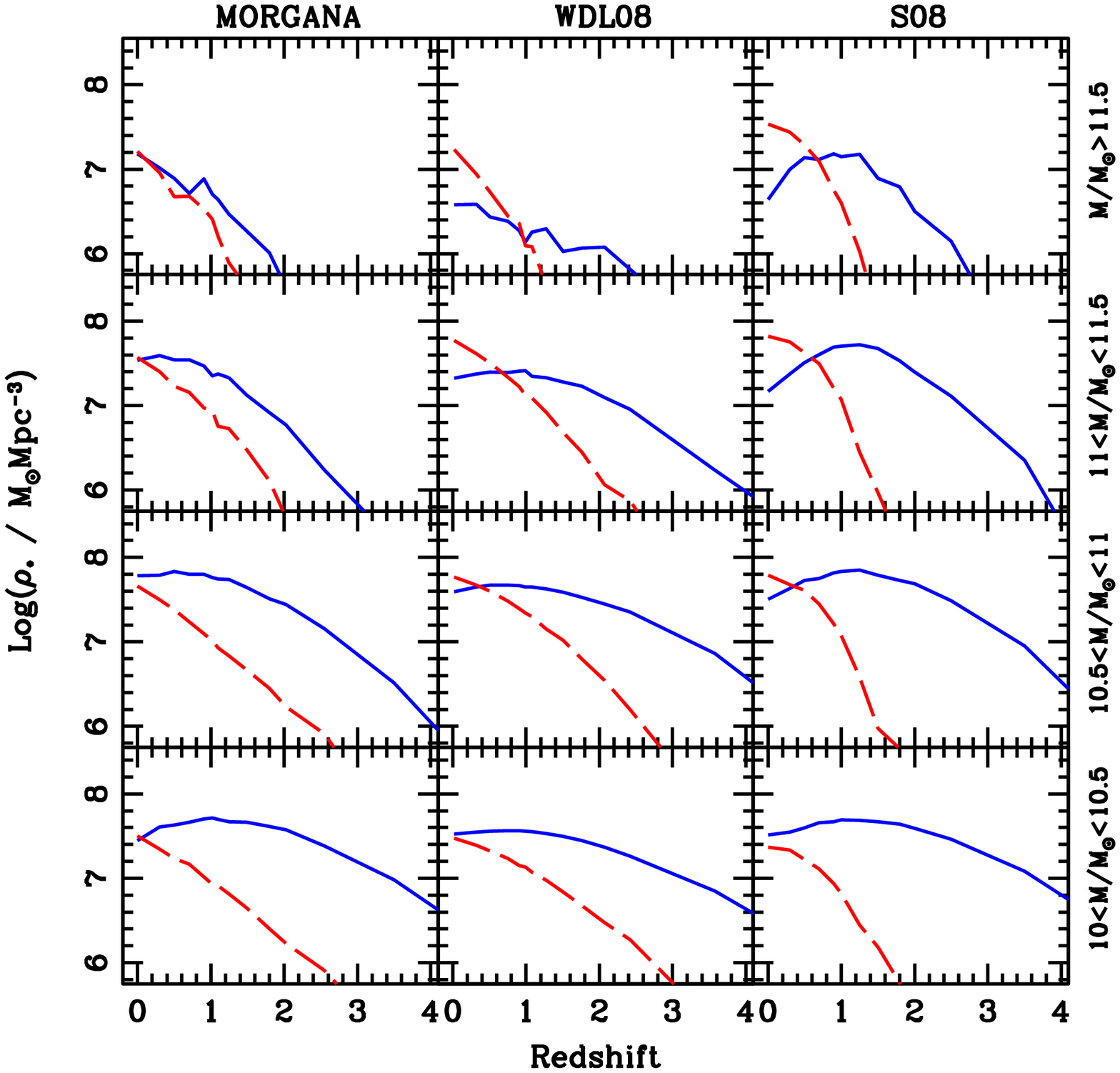} }
\caption{Stellar mass density of active (SSFR$>
  10^{-11}\ {\rm yr}^{-1}$, blue solid lines) and passive (SSFR$<
  10^{-11}\ {\rm yr}^{-1}$, red dashed lines) galaxies in bins of
  galaxy stellar mass, as a function of redshift. Model predictions
  have been convolved with a Gaussian error on $\log M_\star$ with a
  standard deviation of 0.25 dex.}
\label{fig:actpas}
\end{figure}

In Fig.~\ref{fig:actpas}, we show the stellar mass-weighted integrals
of these functions, i.e., the stellar mass density contained in the
active and passive populations, as a function of stellar mass and
redshift. In all models, the stellar mass density (SMD) is dominated
by actively star forming galaxies at high redshift, with the SMD
contributed by passive objects growing rapidly at $z\lesssim 1$. These
results are in qualitative agreement with observational results at
$z\lesssim 1$ \citep[e.g.][]{Bell07}. Observational results at higher 
redshift will soon be available from ongoing and future surveys. 

Up until this point in this Section, we have discussed the model-data
comparison without assessing whether either the predicted or observed behavior
constitutes ``downsizing''. The DS-like differential evolution would be seen as
an earlier accumulation of massive passive galaxies in figure~\ref{fig:2dd},
and as a flattening of the slope of the stellar mass--SFR relation in
figure~\ref{fig:msf_evo} with increasing time. In both figures \ref{fig:2dd}
and \ref{fig:msf_evo} we see a clear downward shift over time of the
star-forming sequence both in the observations and in the models.  Given,
however, the significant discrepancies seen between different datasets and
different SF indicators, and the possible incompleteness of the observations of
low-mass passive galaxies at high redshift, we feel that it is difficult to
claim that there is currently robust evidence for this differential evolution
(DS) in the data in either Figure. Once again, however, the models if anything
show a {\em reverse} downsizing trend, with passive low-mass galaxies appearing
earlier than massive, passive galaxies. In Figure~\ref{fig:csfr}, the signature
of DS would be a more rapid drop, with decreasing redshift, of the SFR density
for more massive galaxies.  Although some previous studies have claimed to see
such an effect \citep[e.g.][]{Juneau05}, the observational compilation that we
have shown here does not show clear evidence for this differential decline. The
one bin in which a markedly sharp decline is seen (the highest mass bin, see
discussion in \citealt{Santini09}) may be affected by cosmic variance.  Once
again, the data appear to be consistent with a constant rate of decline in SFR
density for galaxies of all masses.

\section{Archaeological downsizing}
\label{sec:arch_down}

\begin{figure}
\centerline{
  \includegraphics[width=9cm]{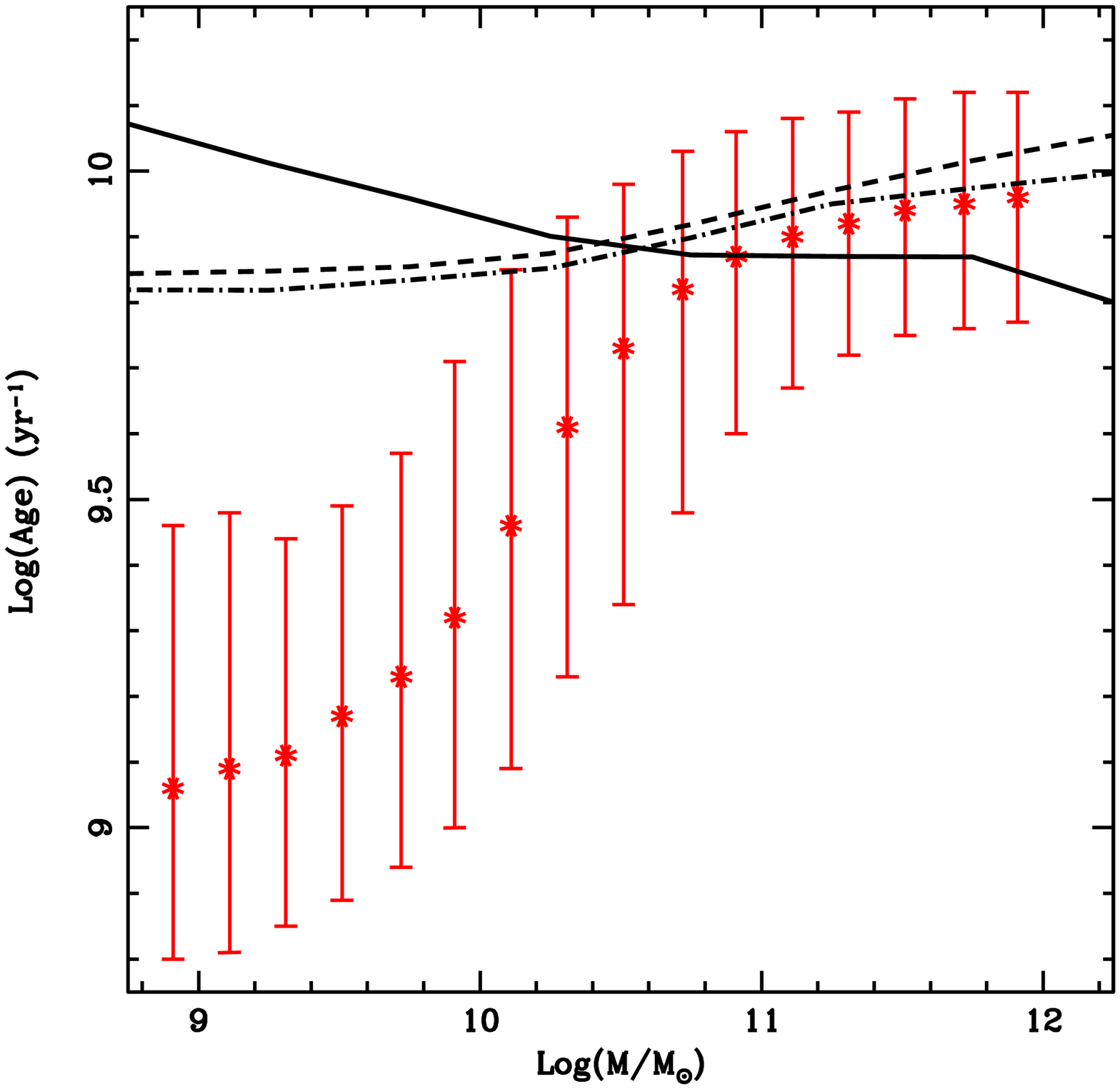} }
\caption{Archaeological DS. Observational constraints on the mean
  mass-weighted stellar age at $z=0$ as a function of stellar mass.
  Solid, dashed and dot-dashed lines refer to the {\mor}, {\mun}, {\som}
  models respectively.  Data from \citet{Gallazzi05}.}
  \label{fig:ages}
\end{figure}

In this section, we focus on the relation between the $z=0$ galaxy
stellar mass and the average age of the stellar population (the
archaeological DS discussed in Section~\ref{intro}).  In
Figure~\ref{fig:ages} we compare the stellar mass-weighted age of the
stellar populations in galaxies as a function of their stellar mass
(at $z=0$) as predicted by the three models with the observational
estimates from \citet{Gallazzi05}. They use high-resolution SDSS
spectra to obtain estimates for the ages and metallicities of $\sim
170,000$ galaxies with $M_\star> 10^9 \msun$. They measure these by
comparing a set of absorption features in the spectra (in particular
the Lick indices and the 4000 {\AA} break) to a grid of synthetic SEDs
covering a wide range of plausible star formation histories and
metallicities.  Both the chosen star formation histories and stellar
population synthesis codes adopted are a likely source of systematic
uncertainty in these estimates.  Moreover, corrections must be made
for in-filling by emission lines in the age-sensitive spectral
features (see \citealt{Gallazzi05} for a complete discussion on how
this correction was applied.

Our results show that the model massive galaxies are old, in agreement
with the observations. However, two out of three models predict only a
mild trend in age from high mass to low mass galaxies, in conflict
with the steeper trend seen in the observational estimates (as already
pointed out by S08). {\mor} behaves like models without AGN feedback,
which produce an inverted trend (in which massive galaxies are younger
than low-mass galaxies); \citet{Croton06} and \citet{DeLucia06} showed
that including the ``radio mode'' AGN feedback makes the massive
galaxies older, improving the agreement with the observed trend. Once
again, it is low-mass galaxies that are discrepant, in the sense that
they form too early and thus have ages that are too old.

The inverted trend predicted by {\mor} is mainly due to two different
physical processes. The younger ages of massive galaxies are related
to the inefficient quenching of cooling flows in massive halos at
$z<1$ \citep[see the discussion in][]{Kimm08}. The resulting higher
level of star formation implies younger ages with respect to {\mun}
and {\som}. The older ages of intermediate-to-low mass sources are
likely due to the enhanced cooling at high-redshift discussed in
\citet{Viola08}, and to the associated enhanced star-formation at
early times.

We note that the observational estimates are closer to being
luminosity-weighted more than stellar mass-weighted ---
\citet{DeLucia06} showed that light-weighted ages show a stronger
trend with stellar mass --- and also that the ages based on absorption
line indices (mainly Balmer lines) tend to actually reflect the age of
the most recent star formation episode, rather than the
luminosity-weighted age \citep{Trager00a,Trager08}. However,
\citet{TragerSomerville08} find that when these observational biases
are modeled by extracting line strengths for the SAM galaxies in the
same way as is done for the observations, this effect cannot fully
account for the discrepancy between the model ages and the observed
ages for low-mass galaxies. 

\begin{figure}
\centerline{
  \includegraphics[width=9cm]{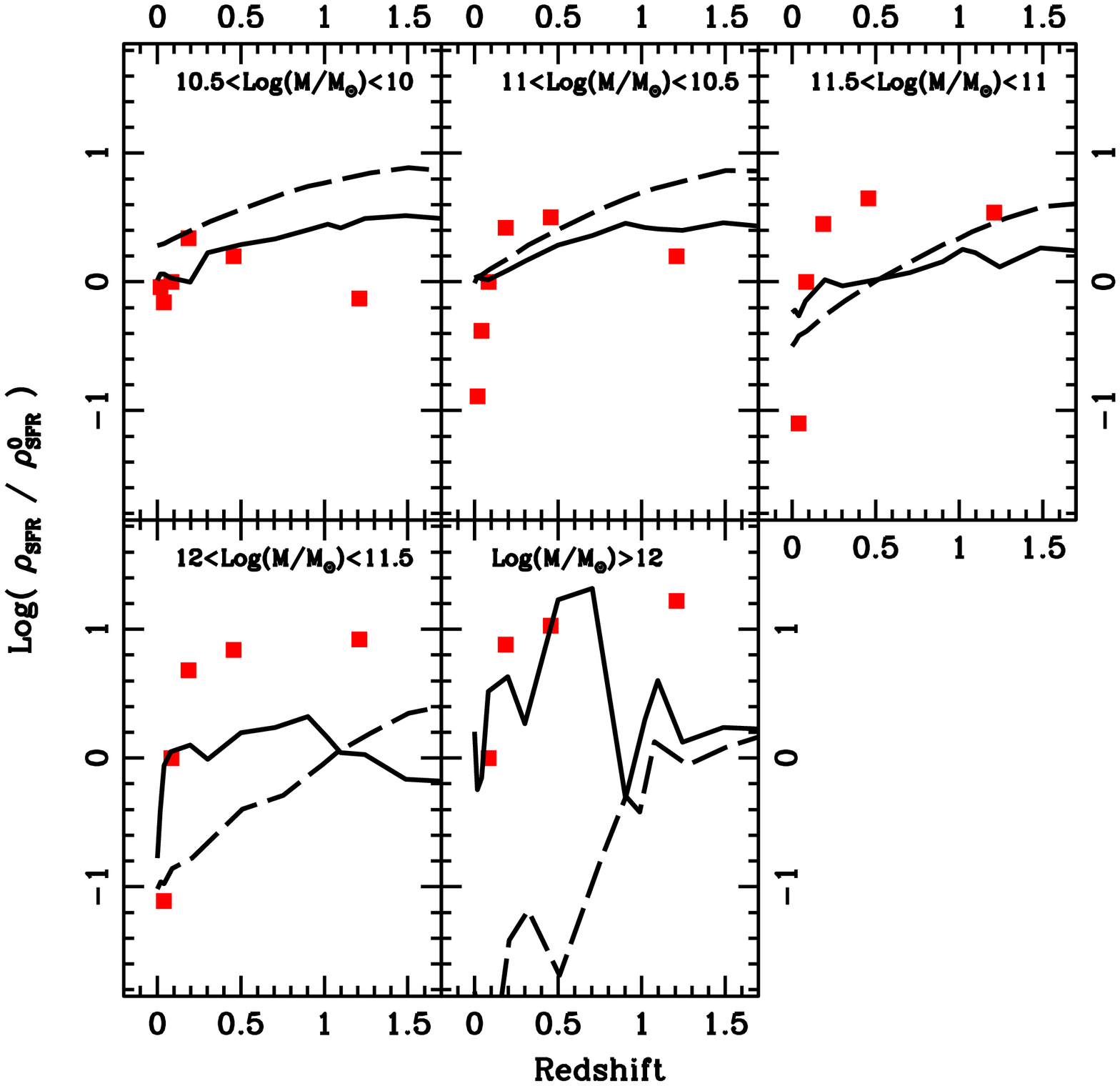}
}
\caption{Cosmic SFR density in bins of $z=0$ stellar
  mass, as a function of redshift. Red squares show observational
  estimates by \citet{Panter07}. Solid and dashed lines refer to the
  {\mor} and {\mun} models respectively (uncertainties on stellar mass
  and SFR have been included).}
\label{fig:archeo}
\end{figure}

One can even go a step further and attempt to extract star formation
histories from galaxy spectra \citep[e.g.][]{Panter07} and to construct
an average SF history for galaxies binned in terms of their stellar
mass at $z=0$. We compare our models with the results of
\citet{Panter07}, who applied the MOPED algorithm to high resolution
($3$ \AA) spectra from the SDSS. This algorithm is similar in spirit
to the SED-fitting we described in sec.~\ref{sec:mass_down}, but it
treats the SFR as a free parameter (defined on an 11-bin grid), thus
allowing for the reconstruction of the star formation history of
galaxies. We stress that these measurements come with numerous
uncertainties. \citet{Panter07} showed that their reconstructed SF
histories depend strongly on the input assumptions. In particular, they
demonstrated that the largest systematics are related to the chosen
spectrophotometric code, stellar population model, the assumed IMF,
the dust attenuation prescription, and the calibration of the observed
spectra.

These observationally derived SF histories are shown in
Fig.~\ref{fig:archeo}, where we plot the cosmic SFR density in bins of
$z=0$ stellar mass as a function of redshift. In each panel we
renormalize both the data and the model predictions to the observed
value at $z=0.0844$ in order to highlight the differences in the
shapes. For technical reasons, we cannot easily extract the star
formation histories for galaxies selected by present-day stellar mass
from the {\som} models. We therefore limit this final comparison to
the two other models. Fig.~\ref{fig:archeo} shows that both SAMs
considered here fail to reproduce the observed trend in detail.  Small
galaxies form too large a fraction of their stars (compared to the
observational estimate) at high redshift. For more massive galaxies,
the SFR density evolution seems fairly well reproduced by {\mor},
while in the {\mun} model too few stars are produced in massive
galaxies at low redshift.

\section{Discussion \& Conclusions}
\label{sec:final}

We have presented a systematic comparison of semi-analytic models
(SAMs) of galaxy formation with observations of local and
high-redshift galaxies that have been claimed to show so-called
``downsizing'' trends.  We had several goals: (i) to re-assess the
robustness of the claims of observed downsizing in the literature,
based on an extensive comparison of different observational data-sets
ii) to see if a consistent picture is painted by the different
observational ``manifestations'' of downsizing iii) to test to what
extent the predictions of hierarchical models of galaxy formation, set
within the $\Lambda$CDM framework, are consistent with these
observational results.

In order to test the general paradigm of galaxy formation within the
hierarchical picture rather than a specific model implementation, we
considered predictions from three independently developed SAMs
({\mun}, {\mor} and {\som}).  We used physical quantities (stellar
masses and SFRs) derived from observations to avoid confusion related
to differences arising from the spectro-photometric codes and dust
models used by the three SAMs. Of course, we cannot avoid these issues
since the observational estimates of stellar masses and SFRs still
depend on stellar population models and contain assumptions about dust
content, metallicity, star formation history, and IMF.

Despite significant differences in the recipes adopted in the three models to
describe the physical processes acting on the baryonic component, the
predictions are remarkably consistent both for the evolution of the stellar
mass and for the star formation history.  This is encouraging, in that it
suggests that our results are relatively robust to the details of the model
assumptions.

We summarize our findings in terms of the three different
manifestations of downsizing that we considered here. We remind the
reader that, in all cases, the signature of downsizing is that massive
galaxies formed (or were assembled) earlier and more rapidly than
lower-mass galaxies.

{\em DS in stellar mass}: (i) We do not see robust evidence for {\it
  differential evolution} of the stellar mass assembly in the
observations, i.e., the data are consistent with an increase in
stellar mass density at the same rate for all stellar mass bins. (ii)
We find that the models roughly reproduce the evolution of the space
density of massive galaxies when their predictions are convolved with
a realistic estimate for the observational error on stellar masses.
At the same time, all models predict almost no evolution in the number
density of galaxies of mass $\sim10^{10}\ \msun$ since $z\sim2$, at
variance with real galaxies whose number density evolves by a factor
of $\sim6$ in the same redshift interval. Put another way, the models
(which are normalized to reproduce the stellar mass function at $z=0$)
{\em overproduce} low-mass galaxies relative to observations at high
redshift ($z\gtrsim0.5$).

{\em DS in SFR}: (i) We find that different estimates of SFR as a
function of stellar mass from different methods show large systematic
offsets as well as differences in slope. Based on the available
observational compilation, we do not see conclusive evidence for {\it
  differential} evolution of the SFR or SFR density for galaxies of
different mass. It may therefore be premature to reach any firm
conclusions about whether these observations in fact show the
signatures of DS. (ii) The models roughly reproduce the increase of
the average SSFR and SFR density of galaxies up to $z\sim4$ though
with a possible systematic underestimate, the weak evolution of the
stellar MF of actively star-forming galaxies and the build-up of the
population of passive galaxies at $z<2$.  However, the mass function
of passive galaxies has a much steeper small-mass-end slope than the
data, and low-mass galaxies are too passive (have too little star
formation) at all probed redshifts.

{\em Archaeological DS}: (i) The data do clearly show the trend of
massive galaxies being older than low-mass galaxies. However, this
trend may arise in part from biases related to the SFR reconstruction
algorithms.  (ii) The {\mun} and {\som} SAMs qualitatively reproduce
the observed trend, in that low-mass galaxies are younger than
high-mass ones.  However, the slope of the mean stellar population age
vs. stellar mass trend is much shallower in the models than in the
data. Some, though probably not all, of this discrepancy may be
related to observational biases in the age estimates.  The SAMs do not
agree well with the detailed star formation histories as a function of
$z=0$ stellar mass extracted from galaxy spectra; low mass galaxies
form too large a fraction of their stars at early times, and high mass
galaxies (at least in the {\mun} models) do not have enough star
formation at late times.

Massive galaxies have long been considered one of the main challenges
for hierarchical models.  The introduction of so-called ``radio mode''
AGN feedback helps keep massive galaxies from forming stars down to
$z=0$, so that red and old massive galaxies are now produced by the
latest generation of SAMs.  We find that when the stellar mass errors
are accounted for \citep{Borch06,Baugh06,Kitzbichler07}, discrepancies
in the number densities of massive galaxies weaken or disappear.  A
number of problems still affect model predictions for the most massive
galaxies: according to the results shown above, their evolution since
$z\sim1$, which is driven by mergers, is marginally inconsistent with
the data (Figure~\ref{fig:smd_evo}). Models may also underestimate the
number of massive galaxies at $z>2$ \citep[see also][]{Marchesini08}.
Depending in part on which observational estimates turn out to be
correct, at least in some of the models the SFR in massive galaxies at
high redshift may be too low.  These residual discrepancies may be
solved by better modeling the known processes: the implementation of
AGN feedback is still extremely crude.  The merger-driven evolution at
$z<1$ may be slowed down by scattering of stars into the diffuse
stellar component of galaxy groups and clusters
\citep{Monaco06,Conroy07,Somerville08}.  Moreover, better modeling of
chemical evolution is needed to address what may be the most severe
challenge for massive galaxies, the chemo-archaeological DS.

At the same time, we find serious discrepancies in the model
predictions for less massive galaxies in the range
$10^9-10^{11}\ \msun$ in stellar mass: they form too early and have
too little ongoing star formation at later times, so their stellar
populations are too old at $z=0$. Their number density is nearly
constant since $z\sim2$, while observations show that it grows in
time. Their SSFR is too low compared with observational data. The
low-mass end slope of the SMF of passive galaxies is too steep, again
indicating an excess of low-mass passive galaxies. Part of this
discrepancy could be due to the over-quenching problem for satellite
galaxies \citep{Weinmann06b,vdBosch08,Kimm08,GilbankBalogh08}, which
is caused by the assumption in all three SAMs that the hot halo is
instantly stripped from satellites as they enter a larger host halo,
thus shutting off any further cooling onto satellite
galaxies. However, as we showed in Figure~\ref{fig:cen_sat}, the
problematic galaxies are predominantly central galaxies in DM haloes
with relatively high circular velocities, $\sim100-200$
km/s. Therefore, mechanisms that only impact satellite galaxies (such
as ram pressure stripping), or that only work on very low mass haloes
(like photo-ionization or, probably, pre-heating) are not viable
solutions to this problem.

The paradox is that we must suppress the formation of low-mass
galaxies in order to fit the low-mass end of the stellar mass function
or the faint end of the luminosity function within the CDM
paradigm. In the three models presented here, as in probably all
$\Lambda$CDM models in the literature, this is currently accomplished
by implementing very strong supernova feedback in low-mass
galaxies. Not only is it unclear that this strong SN feedback is
physically motivated or in agreement with direct observations of winds
in low-mass galaxies, but apparently it does not produce the correct
formation histories for low-mass galaxies. 

Another hint may come from chemical DS: \citet{Maiolino08} \citep[see
  also][]{LoFaro09} showed that the models predict that small galaxies
at high redshift are much more metal-rich than observed galaxies at
these mass scales. This could either indicate that the metals are
efficiently removed from these galaxies, e.g. by winds, or that star
formation (and therefore metal production) is inefficient. 

Thinking of a plausible mechanism that can suppress the formation of
galaxies in small but compact DM haloes at high $z$ is not so easy:
their density is too high and their potential wells are too deep to
suppress star formation with heating from an external UV background,
while massive galactic winds should not destroy galaxies of the same
circular velocity at lower redshift.  Therefore, the discrepancies
discussed above call for a deep re-thinking of the feedback schemes
currently implemented in SAMs.  Alternatively, the problem may be
related to the nature of DM; if this is not completely collisionless,
the density profiles of small DM haloes may be significantly different
from the generally assumed \citet{Navarro96} form, and this would
influence cooling rates and infall times, galaxy sizes and SFRs.

All model predictions discussed in this paper and the data shown in
Fig.~2 are available in electronic format upon request.

\section*{Acknowledgments}
We are grateful to Eric Bell and Anna Gallazzi for discussion and
careful explanation of their data, to Frank van den Bosch, Maurilio
Pannella and Nicola Menci for enlighting discussions, to Adriano
Fontana, Andrea Grazian, Sara Salimbeni for help in understanding and
extracting information from the GOODS-MUSIC catalogue, to Danilo
Marchesini for sharing his data before publication, and to Kai Noeske
for providing his data in electronic form and for very useful
discussions about star formation indicators. FF and GDL acknowledge
hospitality at the Kavli Institute for Theoretical Physics in Santa
Barbara. This research was supported in part by the National Science
Foundation under Grant No. NSF PHY05-51164. We thank the anonymous
referee for suggestions that helped to improve this paper.

\bibliographystyle{mn2e}
\bibliography{fontanot}

\begin{thebibliography}{}

\bibitem[\protect\citeauthoryear{{Ando}, {Ohta}, {Iwata}, {Akiyama}, {Aoki} \&
  {Tamura}}{{Ando} et~al.}{2007}]{Ando07}
{Ando} M.,  {Ohta} K.,  {Iwata} I.,  {Akiyama} M.,  {Aoki} K.,    {Tamura} N.,
  2007, \pasj, 59, 717

\bibitem[\protect\citeauthoryear{{Bauer}, {Drory}, {Hill} \& {Feulner}}{{Bauer}
  et~al.}{2005}]{Bauer05}
{Bauer} A.~E.,  {Drory} N.,  {Hill} G.~J.,    {Feulner} G.,  2005, \apjl, 621,
  L89

\bibitem[\protect\citeauthoryear{{Baugh}}{{Baugh}}{2006}]{Baugh06}
{Baugh} C.~M.,  2006, Reports of Progress in Physics, 69, 3101

\bibitem[\protect\citeauthoryear{{Baugh}, {Cole} \& {Frenk}}{{Baugh}
  et~al.}{1996}]{BaughColeFrenk96}
{Baugh} C.~M.,  {Cole} S.,    {Frenk} C.~S.,  1996, \mnras, 283, 1361

\bibitem[\protect\citeauthoryear{{Bell}, {McIntosh}, {Katz} \&
  {Weinberg}}{{Bell} et~al.}{2003}]{Bell03}
{Bell} E.~F.,  {McIntosh} D.~H.,  {Katz} N.,    {Weinberg} M.~D.,  2003, \apjs,
  149, 289

\bibitem[\protect\citeauthoryear{{Bell}, {Zheng}, {Papovich}, {Borch}, {Wolf}
  \& {Meisenheimer}}{{Bell} et~al.}{2007}]{Bell07}
{Bell} E.~F.,  {Zheng} X.~Z.,  {Papovich} C.,  {Borch} A.,  {Wolf} C.,
  {Meisenheimer} K.,  2007, \apj, 663, 834

\bibitem[\protect\citeauthoryear{{Borch}, {Meisenheimer}, {Bell}, {Rix},
  {Wolf}, {Dye}, {Kleinheinrich}, {Kovacs} \& {Wisotzki}}{{Borch}
  et~al.}{2006}]{Borch06}
{Borch} A.,  {Meisenheimer} K.,  {Bell} E.~F.,  {Rix} H.-W.,  {Wolf} C.,  {Dye}
  S.,  {Kleinheinrich} M.,  {Kovacs} Z.,    {Wisotzki} L.,  2006, \aap, 453,
  869

\bibitem[\protect\citeauthoryear{{Boselli}, {Gavazzi}, {Donas} \&
  {Scodeggio}}{{Boselli} et~al.}{2001}]{Boselli01}
{Boselli} A.,  {Gavazzi} G.,  {Donas} J.,    {Scodeggio} M.,  2001, \aj, 121,
  753

\bibitem[\protect\citeauthoryear{{Bower}, {Benson}, {Malbon}, {Helly}, {Frenk},
  {Baugh}, {Cole} \& {Lacey}}{{Bower} et~al.}{2006}]{Bower06}
{Bower} R.~G.,  {Benson} A.~J.,  {Malbon} R.,  {Helly} J.~C.,  {Frenk} C.~S.,
  {Baugh} C.~M.,  {Cole} S.,    {Lacey} C.~G.,  2006, \mnras, 370, 645

\bibitem[\protect\citeauthoryear{{Boylan-Kolchin}, {Ma} \&
  {Quataert}}{{Boylan-Kolchin} et~al.}{2008}]{BoylanKolchin08}
{Boylan-Kolchin} M.,  {Ma} C.-P.,    {Quataert} E.,  2008, \mnras, 383, 93

\bibitem[\protect\citeauthoryear{{Brinchmann}, {Charlot}, {White}, {Tremonti},
  {Kauffmann}, {Heckman} \& {Brinkmann}}{{Brinchmann}
  et~al.}{2004}]{Brinchmann04}
{Brinchmann} J.,  {Charlot} S.,  {White} S.~D.~M.,  {Tremonti} C.,  {Kauffmann}
  G.,  {Heckman} T.,    {Brinkmann} J.,  2004, \mnras, 351, 1151

\bibitem[\protect\citeauthoryear{{Bundy}, {Ellis}, {Conselice}, {Taylor},
  {Cooper}, {Willmer}, {Weiner}, {Coil}, {Noeske} \& {Eisenhardt}}{{Bundy}
  et~al.}{2006}]{Bundy06}
{Bundy} K.,  {Ellis} R.~S.,  {Conselice} C.~J.,  {Taylor} J.~E.,  {Cooper}
  M.~C.,  {Willmer} C.~N.~A.,  {Weiner} B.~J.,  {Coil} A.~L.,  {Noeske} K.~G.,
    {Eisenhardt} P.~R.~M.,  2006, \apj, 651, 120

\bibitem[\protect\citeauthoryear{{Carollo}, {Danziger} \& {Buson}}{{Carollo}
  et~al.}{1993}]{Carollo93}
{Carollo} C.~M.,  {Danziger} I.~J.,    {Buson} L.,  1993, \mnras, 265, 553

\bibitem[\protect\citeauthoryear{{Chabrier}}{{Chabrier}}{2003}]{Chabrier03}
{Chabrier} G.,  2003, \apjl, 586, L133

\bibitem[\protect\citeauthoryear{{Chen}, {Wild}, {Kauffmann}, {Blaizot},
  {Davis}, {Noeske}, {Wang} \& {Willmer}}{{Chen} et~al.}{2009}]{Chen09}
{Chen} Y.-M.,  {Wild} V.,  {Kauffmann} G.,  {Blaizot} J.,  {Davis} M.,
  {Noeske} K.,  {Wang} J.-M.,    {Willmer} C.,  2009, \mnras, 393, 406

\bibitem[\protect\citeauthoryear{{Cimatti}, {Daddi} \& {Renzini}}{{Cimatti}
  et~al.}{2006}]{Cimatti06}
{Cimatti} A.,  {Daddi} E.,    {Renzini} A.,  2006, \aap, 453, L29

\bibitem[\protect\citeauthoryear{{Cirasuolo}, {McLure}, {Dunlop}, {Almaini},
  {Foucaud} \& {Simpson}}{{Cirasuolo} et~al.}{2008}]{Cirasuolo08}
{Cirasuolo} M.,  {McLure} R.~J.,  {Dunlop} J.~S.,  {Almaini} O.,  {Foucaud} S.,
     {Simpson} C.,  2008, ArXiv e-prints, 804

\bibitem[\protect\citeauthoryear{{Cole}, {Norberg}, {Baugh}, {Frenk},
  {Bland-Hawthorn}, {Bridges}, {Cannon} \& {Colless}}{{Cole}
  et~al.}{2001}]{Cole01}
{Cole} S.,  {Norberg} P.,  {Baugh} C.~M.,  {Frenk} C.~S.,  {Bland-Hawthorn} J.,
   {Bridges} T.,  {Cannon} R.,    {Colless} M. e.~a.,  2001, \mnras, 326, 255

\bibitem[\protect\citeauthoryear{{Conroy}, {Wechsler} \& {Kravtsov}}{{Conroy}
  et~al.}{2007}]{Conroy07}
{Conroy} C.,  {Wechsler} R.~H.,    {Kravtsov} A.~V.,  2007, \apj, 668, 826

\bibitem[\protect\citeauthoryear{{Conselice}, {Bundy}, {Trujillo}, {Coil},
  {Eisenhardt}, {Ellis}, {Georgakakis}, {Huang}, {Lotz}, {Nandra}, {Newman},
  {Papovich}, {Weiner} \& {Willmer}}{{Conselice} et~al.}{2007}]{Conselice07}
{Conselice} C.~J.,  {Bundy} K.,  {Trujillo} I.,  {Coil} A.,  {Eisenhardt} P.,
  {Ellis} R.~S.,  {Georgakakis} A.,  {Huang} J.,  {Lotz} J.,  {Nandra} K.,
  {Newman} J.,  {Papovich} C.,  {Weiner} B.,    {Willmer} C.,  2007, \mnras,
  381, 962

\bibitem[\protect\citeauthoryear{{Cowie} \& {Barger}}{{Cowie} \&
  {Barger}}{2008}]{Cowie08}
{Cowie} L.~L.,  {Barger} A.~J.,  2008, \apj, 686, 72

\bibitem[\protect\citeauthoryear{{Cowie}, {Songaila}, {Hu} \& {Cohen}}{{Cowie}
  et~al.}{1996}]{Cowie96}
{Cowie} L.~L.,  {Songaila} A.,  {Hu} E.~M.,    {Cohen} J.~G.,  1996, \aj, 112,
  839

\bibitem[\protect\citeauthoryear{{Cristiani}, {Alexander}, {Bauer}, {Brandt},
  {Chatzichristou}, {Fontanot}, {Grazian}, {Koekemoer}, {Lucas}, {Monaco},
  {Nonino}, {Padovani}, {Stern}, {Tozzi}, {Treister}, {Urry} \&
  {Vanzella}}{{Cristiani} et~al.}{2004}]{Cristiani04}
{Cristiani} S.,  {Alexander} D.~M.,  {Bauer} F.,  {Brandt} W.~N.,
  {Chatzichristou} E.~T.,  {Fontanot} F.,  {Grazian} A.,  {Koekemoer} A.,
  {Lucas} R.~A.,  {Monaco} P.,  {Nonino} M.,  {Padovani} P.,  {Stern} D.,
  {Tozzi} P.,  {Treister} E.,  {Urry} C.~M.,    {Vanzella} E.,  2004, \apjl,
  600, L119

\bibitem[\protect\citeauthoryear{{Croton}, {Springel}, {White}, {De Lucia},
  {Frenk}, {Gao}, {Jenkins}, {Kauffmann}, {Navarro} \& {Yoshida}}{{Croton}
  et~al.}{2006}]{Croton06}
{Croton} D.~J.,  {Springel} V.,  {White} S.~D.~M.,  {De Lucia} G.,  {Frenk}
  C.~S.,  {Gao} L.,  {Jenkins} A.,  {Kauffmann} G.,  {Navarro} J.~F.,
  {Yoshida} N.,  2006, \mnras, 365, 11

\bibitem[\protect\citeauthoryear{{Daddi}, {Dickinson}, {Morrison}, {Chary},
  {Cimatti}, {Elbaz}, {Frayer}, {Renzini}, {Pope}, {Alexander}, {Bauer},
  {Giavalisco}, {Huynh}, {Kurk} \& {Mignoli}}{{Daddi} et~al.}{2007}]{Daddi07}
{Daddi} E.,  {Dickinson} M.,  {Morrison} G.,  {Chary} R.,  {Cimatti} A.,
  {Elbaz} D.,  {Frayer} D.,  {Renzini} A.,  {Pope} A.,  {Alexander} D.~M.,
  {Bauer} F.~E.,  {Giavalisco} M.,  {Huynh} M.,  {Kurk} J.,    {Mignoli} M.,
  2007, \apj, 670, 156

\bibitem[\protect\citeauthoryear{{Davies}, {Sadler} \& {Peletier}}{{Davies}
  et~al.}{1993}]{Davies93}
{Davies} R.~L.,  {Sadler} E.~M.,    {Peletier} R.~F.,  1993, \mnras, 262, 650

\bibitem[\protect\citeauthoryear{{De Lucia} \& {Blaizot}}{{De Lucia} \&
  {Blaizot}}{2007}]{DeLucia07b}
{De Lucia} G.,  {Blaizot} J.,  2007, \mnras, 375, 2

\bibitem[\protect\citeauthoryear{{De Lucia} \& {Helmi}}{{De Lucia} \&
  {Helmi}}{2008}]{DeLuciaHelmi08}
{De Lucia} G.,  {Helmi} A.,  2008, \mnras, 391, 14

\bibitem[\protect\citeauthoryear{{De Lucia}, {Kauffmann}, {Springel}, {White},
  {Lanzoni}, {Stoehr}, {Tormen} \& {Yoshida}}{{De Lucia}
  et~al.}{2004}]{DeLucia04a}
{De Lucia} G.,  {Kauffmann} G.,  {Springel} V.,  {White} S.~D.~M.,  {Lanzoni}
  B.,  {Stoehr} F.,  {Tormen} G.,    {Yoshida} N.,  2004, \mnras, 348, 333

\bibitem[\protect\citeauthoryear{{De Lucia}, {Poggianti},
  {Arag{\'o}n-Salamanca}, {Clowe}, {Halliday}, {Jablonka}, {Milvang-Jensen},
  {Pell{\'o}}, {Poirier}, {Rudnick}, {Saglia}, {Simard} \& {White}}{{De Lucia}
  et~al.}{2004}]{DeLucia04c}
{De Lucia} G.,  {Poggianti} B.~M.,  {Arag{\'o}n-Salamanca} A.,  {Clowe} D.,
  {Halliday} C.,  {Jablonka} P.,  {Milvang-Jensen} B.,  {Pell{\'o}} R.,
  {Poirier} S.,  {Rudnick} G.,  {Saglia} R.,  {Simard} L.,    {White} S.~D.~M.,
   2004, \apjl, 610, L77

\bibitem[\protect\citeauthoryear{{De Lucia}, {Poggianti},
  {Arag{\'o}n-Salamanca}, {White}, {Zaritsky}, {Clowe}, {Halliday}, {Jablonka},
  {von der Linden}, {Milvang-Jensen}, {Pell{\'o}}, {Rudnick}, {Saglia} \&
  {Simard}}{{De Lucia} et~al.}{2007}]{DeLucia07a}
{De Lucia} G.,  {Poggianti} B.~M.,  {Arag{\'o}n-Salamanca} A.,  {White}
  S.~D.~M.,  {Zaritsky} D.,  {Clowe} D.,  {Halliday} C.,  {Jablonka} P.,  {von
  der Linden} A.,  {Milvang-Jensen} B.,  {Pell{\'o}} R.,  {Rudnick} G.,
  {Saglia} R.~P.,    {Simard} L.,  2007, \mnras, 374, 809

\bibitem[\protect\citeauthoryear{{De Lucia}, {Springel}, {White}, {Croton} \&
  {Kauffmann}}{{De Lucia} et~al.}{2006}]{DeLucia06}
{De Lucia} G.,  {Springel} V.,  {White} S.~D.~M.,  {Croton} D.,    {Kauffmann}
  G.,  2006, \mnras, 366, 499

\bibitem[\protect\citeauthoryear{{Drory} \& {Alvarez}}{{Drory} \&
  {Alvarez}}{2008}]{Drory08}
{Drory} N.,  {Alvarez} M.,  2008, \apj, 680, 41

\bibitem[\protect\citeauthoryear{{Drory}, {Bender}, {Feulner}, {Hopp},
  {Maraston}, {Snigula} \& {Hill}}{{Drory} et~al.}{2004}]{Drory04}
{Drory} N.,  {Bender} R.,  {Feulner} G.,  {Hopp} U.,  {Maraston} C.,  {Snigula}
  J.,    {Hill} G.~J.,  2004, \apj, 608, 742

\bibitem[\protect\citeauthoryear{{Drory}, {Salvato}, {Gabasch}, {Bender},
  {Hopp}, {Feulner} \& {Pannella}}{{Drory} et~al.}{2005}]{Drory05}
{Drory} N.,  {Salvato} M.,  {Gabasch} A.,  {Bender} R.,  {Hopp} U.,  {Feulner}
  G.,    {Pannella} M.,  2005, \apjl, 619, L131

\bibitem[\protect\citeauthoryear{{Dunne}, {Ivison}, {Maddox}, {Cirasuolo},
  {Mortier}, {Foucaud}, {Ibar}, {Almaini}, {Simpson} \& {McLure}}{{Dunne}
  et~al.}{2008}]{Dunne08}
{Dunne} L.,  {Ivison} R.~J.,  {Maddox} S.,  {Cirasuolo} M.,  {Mortier} A.~M.,
  {Foucaud} S.,  {Ibar} E.,  {Almaini} O.,  {Simpson} C.,    {McLure} R.,
  2008, ArXiv e-prints, 808

\bibitem[\protect\citeauthoryear{{Elbaz}, {Daddi}, {Le Borgne}, {Dickinson},
  {Alexander}, {Chary}, {Starck}, {Brandt}, {Kitzbichler}, {MacDonald},
  {Nonino}, {Popesso}, {Stern} \& {Vanzella}}{{Elbaz} et~al.}{2007}]{Elbaz07}
{Elbaz} D.,  {Daddi} E.,  {Le Borgne} D.,  {Dickinson} M.,  {Alexander} D.~M.,
  {Chary} R.-R.,  {Starck} J.-L.,  {Brandt} W.~N.,  {Kitzbichler} M.,
  {MacDonald} E.,  {Nonino} M.,  {Popesso} P.,  {Stern} D.,    {Vanzella} E.,
  2007, \aap, 468, 33

\bibitem[\protect\citeauthoryear{{Erb}, {Shapley}, {Pettini}, {Steidel},
  {Reddy} \& {Adelberger}}{{Erb} et~al.}{2006}]{Erb06}
{Erb} D.~K.,  {Shapley} A.~E.,  {Pettini} M.,  {Steidel} C.~C.,  {Reddy} N.~A.,
     {Adelberger} K.~L.,  2006, \apj, 644, 813

\bibitem[\protect\citeauthoryear{{Faber}, {Worthey} \& {Gonzales}}{{Faber}
  et~al.}{1992}]{Faber92}
{Faber} S.~M.,  {Worthey} G.,    {Gonzales} J.~J.,  1992, in {Barbuy} B.,
  {Renzini} A.,  eds, The Stellar Populations of Galaxies Vol.~149 of IAU
  Symposium, {Absorption-Line Spectra of Elliptical Galaxies and Their Relation
  to Elliptical Formation}.
pp 255--+

\bibitem[\protect\citeauthoryear{{Feulner}, {Gabasch}, {Salvato}, {Drory},
  {Hopp} \& {Bender}}{{Feulner} et~al.}{2005}]{Feulner05}
{Feulner} G.,  {Gabasch} A.,  {Salvato} M.,  {Drory} N.,  {Hopp} U.,
  {Bender} R.,  2005, \apjl, 633, L9

\bibitem[\protect\citeauthoryear{{Fontana}, {Pozzetti}, {Donnarumma},
  {Renzini}, {Cimatti}, {Zamorani}, {Menci}, {Daddi}, {Giallongo}, {Mignoli},
  {Perna}, {Salimbeni}, {Saracco}, {Broadhurst}, {Cristiani}, {D'Odorico} \&
  {Gilmozzi}}{{Fontana} et~al.}{2004}]{Fontana04}
{Fontana} A.,  {Pozzetti} L.,  {Donnarumma} I.,  {Renzini} A.,  {Cimatti} A.,
  {Zamorani} G.,  {Menci} N.,  {Daddi} E.,  {Giallongo} E.,  {Mignoli} M.,
  {Perna} C.,  {Salimbeni} S.,  {Saracco} P.,  {Broadhurst} T.,  {Cristiani}
  S.,  {D'Odorico} S.,    {Gilmozzi} R.,  2004, \aap, 424, 23

\bibitem[\protect\citeauthoryear{{Fontana}, {Salimbeni}, {Grazian},
  {Giallongo}, {Pentericci}, {Nonino}, {Fontanot}, {Menci}, {Monaco},
  {Cristiani}, {Vanzella}, {de Santis} \& {Gallozzi}}{{Fontana}
  et~al.}{2006}]{Fontana06}
{Fontana} A.,  {Salimbeni} S.,  {Grazian} A.,  {Giallongo} E.,  {Pentericci}
  L.,  {Nonino} M.,  {Fontanot} F.,  {Menci} N.,  {Monaco} P.,  {Cristiani} S.,
   {Vanzella} E.,  {de Santis} C.,    {Gallozzi} S.,  2006, \aap, 459, 745

\bibitem[\protect\citeauthoryear{{Fontanot}, {Cristiani}, {Monaco}, {Nonino},
  {Vanzella}, {Brandt}, {Grazian} \& {Mao}}{{Fontanot}
  et~al.}{2007}]{Fontanot07a}
{Fontanot} F.,  {Cristiani} S.,  {Monaco} P.,  {Nonino} M.,  {Vanzella} E.,
  {Brandt} W.~N.,  {Grazian} A.,    {Mao} J.,  2007, \aap, 461, 39

\bibitem[\protect\citeauthoryear{{Fontanot}, {Monaco}, {Cristiani} \&
  {Tozzi}}{{Fontanot} et~al.}{2006}]{Fontanot06}
{Fontanot} F.,  {Monaco} P.,  {Cristiani} S.,    {Tozzi} P.,  2006, \mnras,
  373, 1173

\bibitem[\protect\citeauthoryear{{Fontanot}, {Monaco}, {Silva} \&
  {Grazian}}{{Fontanot} et~al.}{2007}]{Fontanot07b}
{Fontanot} F.,  {Monaco} P.,  {Silva} L.,    {Grazian} A.,  2007, \mnras, 382,
  903

\bibitem[\protect\citeauthoryear{{Fontanot}, {Somerville}, {Silva}, {Monaco} \&
  {Skibba}}{{Fontanot} et~al.}{2009}]{Fontanot08}
{Fontanot} F.,  {Somerville} R.~S.,  {Silva} L.,  {Monaco} P.,    {Skibba} R.,
  2009, \mnras, 392, 553

\bibitem[\protect\citeauthoryear{{Gallazzi}, {Charlot}, {Brinchmann}, {White}
  \& {Tremonti}}{{Gallazzi} et~al.}{2005}]{Gallazzi05}
{Gallazzi} A.,  {Charlot} S.,  {Brinchmann} J.,  {White} S.~D.~M.,
  {Tremonti} C.~A.,  2005, \mnras, 362, 41

\bibitem[\protect\citeauthoryear{{Gao}, {White}, {Jenkins}, {Stoehr} \&
  {Springel}}{{Gao} et~al.}{2004}]{Gao04}
{Gao} L.,  {White} S.~D.~M.,  {Jenkins} A.,  {Stoehr} F.,    {Springel} V.,
  2004, \mnras, 355, 819

\bibitem[\protect\citeauthoryear{{Gilbank} \& {Balogh}}{{Gilbank} \&
  {Balogh}}{2008}]{GilbankBalogh08}
{Gilbank} D.~G.,  {Balogh} M.~L.,  2008, \mnras, 385, L116

\bibitem[\protect\citeauthoryear{{Gilbank}, {Yee}, {Ellingson}, {Gladders},
  {Loh}, {Barrientos} \& {Barkhouse}}{{Gilbank} et~al.}{2008}]{Gilbank08}
{Gilbank} D.~G.,  {Yee} H.~K.~C.,  {Ellingson} E.,  {Gladders} M.~D.,  {Loh}
  Y.-S.,  {Barrientos} L.~F.,    {Barkhouse} W.~A.,  2008, \apj, 673, 742

\bibitem[\protect\citeauthoryear{{Granato}, {Silva}, {Monaco}, {Panuzzo},
  {Salucci}, {De Zotti} \& {Danese}}{{Granato} et~al.}{2001}]{Granato01}
{Granato} G.~L.,  {Silva} L.,  {Monaco} P.,  {Panuzzo} P.,  {Salucci} P.,  {De
  Zotti} G.,    {Danese} L.,  2001, \mnras, 324, 757

\bibitem[\protect\citeauthoryear{{Grazian}, {Fontana}, {de Santis}, {Nonino},
  {Salimbeni}, {Giallongo}, {Cristiani}, {Gallozzi} \& {Vanzella}}{{Grazian}
  et~al.}{2006}]{Grazian06}
{Grazian} A.,  {Fontana} A.,  {de Santis} C.,  {Nonino} M.,  {Salimbeni} S.,
  {Giallongo} E.,  {Cristiani} S.,  {Gallozzi} S.,    {Vanzella} E.,  2006,
  \aap, 449, 951

\bibitem[\protect\citeauthoryear{{Hasinger}, {Miyaji} \& {Schmidt}}{{Hasinger}
  et~al.}{2005}]{Hasinger05}
{Hasinger} G.,  {Miyaji} T.,    {Schmidt} M.,  2005, \aap, 441, 417

\bibitem[\protect\citeauthoryear{{Heavens}, {Panter}, {Jimenez} \&
  {Dunlop}}{{Heavens} et~al.}{2004}]{Heavens04}
{Heavens} A.,  {Panter} B.,  {Jimenez} R.,    {Dunlop} J.,  2004, \nat, 428,
  625

\bibitem[\protect\citeauthoryear{{Juneau}, {Glazebrook}, {Crampton},
  {McCarthy}, {Savaglio}, {Abraham}, {Carlberg}, {Chen}, {Le Borgne}, {Marzke},
  {Roth}, {J{\o}rgensen}, {Hook} \& {Murowinski}}{{Juneau}
  et~al.}{2005}]{Juneau05}
{Juneau} S.,  {Glazebrook} K.,  {Crampton} D.,  {McCarthy} P.~J.,  {Savaglio}
  S.,  {Abraham} R.,  {Carlberg} R.~G.,  {Chen} H.-W.,  {Le Borgne} D.,
  {Marzke} R.~O.,  {Roth} K.,  {J{\o}rgensen} I.,  {Hook} I.,    {Murowinski}
  R.,  2005, \apjl, 619, L135

\bibitem[\protect\citeauthoryear{{Kauffmann}}{{Kauffmann}}{1996}]{Kauffmann96}
{Kauffmann} G.,  1996, \mnras, 281, 487

\bibitem[\protect\citeauthoryear{{Kewley} \& {Ellison}}{{Kewley} \&
  {Ellison}}{2008}]{Kewley08}
{Kewley} L.~J.,  {Ellison} S.~L.,  2008, \apj, 681, 1183

\bibitem[\protect\citeauthoryear{{Kimm}, {Somerville}, {Yi}, {van den Bosch},
  {Salim}, {Fontanot}, {Monaco}, {Mo}, {Pasquali}, {Rich} \& {Yang}}{{Kimm}
  et~al.}{2008}]{Kimm08}
{Kimm} T.,  {Somerville} R.~S.,  {Yi} S.~K.,  {van den Bosch} F.~C.,  {Salim}
  S.,  {Fontanot} F.,  {Monaco} P.,  {Mo} H.,  {Pasquali} A.,  {Rich} R.~M.,
  {Yang} X.,  2008, ArXiv e-prints

\bibitem[\protect\citeauthoryear{{Kitzbichler} \& {White}}{{Kitzbichler} \&
  {White}}{2007}]{Kitzbichler07}
{Kitzbichler} M.~G.,  {White} S.~D.~M.,  2007, \mnras, 376, 2

\bibitem[\protect\citeauthoryear{{Kodama}, {Yamada}, {Akiyama}, {Aoki}, {Doi},
  {Furusawa}, {Fuse} \& {Imanishi}}{{Kodama} et~al.}{2004}]{Kodama04}
{Kodama} T.,  {Yamada} T.,  {Akiyama} M.,  {Aoki} K.,  {Doi} M.,  {Furusawa}
  H.,  {Fuse} T.,    {Imanishi} M. e.~a.,  2004, \mnras, 350, 1005

\bibitem[\protect\citeauthoryear{{Komatsu}, {Dunkley}, {Nolta}, {Bennett},
  {Gold}, {Hinshaw}, {Jarosik}, {Larson}, {Limon}, {Page}, {Spergel},
  {Halpern}, {Hill}, {Kogut}, {Meyer}, {Tucker}, {Weiland}, {Wollack} \&
  {Wright}}{{Komatsu} et~al.}{2009}]{Komatsu09}
{Komatsu} E.,  {Dunkley} J.,  {Nolta} M.~R.,  {Bennett} C.~L.,  {Gold} B.,
  {Hinshaw} G.,  {Jarosik} N.,  {Larson} D.,  {Limon} M.,  {Page} L.,
  {Spergel} D.~N.,  {Halpern} M.,  {Hill} R.~S.,  {Kogut} A.,  {Meyer} S.~S.,
  {Tucker} G.~S.,  {Weiland} J.~L.,  {Wollack} E.,    {Wright} E.~L.,  2009,
  \apjs, 180, 330

\bibitem[\protect\citeauthoryear{{Kuntschner}, {Lucey}, {Smith}, {Hudson} \&
  {Davies}}{{Kuntschner} et~al.}{2001}]{Kuntschner01}
{Kuntschner} H.,  {Lucey} J.~R.,  {Smith} R.~J.,  {Hudson} M.~J.,    {Davies}
  R.~L.,  2001, \mnras, 323, 615

\bibitem[\protect\citeauthoryear{{La Franca}, {Fiore}, {Comastri}, {Perola},
  {Sacchi}, {Brusa}, {Cocchia}, {Feruglio}, {Matt}, {Vignali}, {Carangelo},
  {Ciliegi}, {Lamastra}, {Maiolino}, {Mignoli}, {Molendi} \& {Puccetti}}{{La
  Franca} et~al.}{2005}]{LaFranca05}
{La Franca} F.,  {Fiore} F.,  {Comastri} A.,  {Perola} G.~C.,  {Sacchi} N.,
  {Brusa} M.,  {Cocchia} F.,  {Feruglio} C.,  {Matt} G.,  {Vignali} C.,
  {Carangelo} N.,  {Ciliegi} P.,  {Lamastra} A.,  {Maiolino} R.,  {Mignoli} M.,
   {Molendi} S.,    {Puccetti} S.,  2005, \apj, 635, 864

\bibitem[\protect\citeauthoryear{{Lee}, {Idzi}, {Ferguson}, {Somerville},
  {Wiklind} \& {Giavalisco}}{{Lee} et~al.}{2008}]{Lee09}
{Lee} S.-K.,  {Idzi} R.,  {Ferguson} H.~C.,  {Somerville} R.~S.,  {Wiklind} T.,
     {Giavalisco} M.,  2008, ArXiv e-prints

\bibitem[\protect\citeauthoryear{{Li}, {Mo} \& {Gao}}{{Li} et~al.}{2008}]{Li08}
{Li} Y.,  {Mo} H.~J.,    {Gao} L.,  2008, \mnras, 389, 1419

\bibitem[\protect\citeauthoryear{{Lo Faro}}{{Lo Faro}}{2009}]{LoFaro09}
{Lo Faro} B. e.~a.,  2009, submitted to \mnras

\bibitem[\protect\citeauthoryear{{Maiolino}, {Nagao}, {Grazian}, {Cocchia},
  {Marconi}, {Mannucci}, {Cimatti} \& {Pipino}}{{Maiolino}
  et~al.}{2008}]{Maiolino08}
{Maiolino} R.,  {Nagao} T.,  {Grazian} A.,  {Cocchia} F.,  {Marconi} A.,
  {Mannucci} F.,  {Cimatti} A.,    {Pipino} A. e.~a.,  2008, \aap, 488, 463

\bibitem[\protect\citeauthoryear{{Maraston}, {Daddi}, {Renzini}, {Cimatti},
  {Dickinson}, {Papovich}, {Pasquali} \& {Pirzkal}}{{Maraston}
  et~al.}{2006}]{Maraston06}
{Maraston} C.,  {Daddi} E.,  {Renzini} A.,  {Cimatti} A.,  {Dickinson} M.,
  {Papovich} C.,  {Pasquali} A.,    {Pirzkal} N.,  2006, \apj, 652, 85

\bibitem[\protect\citeauthoryear{{Marchesini}, {van Dokkum}, {Forster
  Schreiber}, {Franx}, {Labbe'} \& {Wuyts}}{{Marchesini}
  et~al.}{2008}]{Marchesini08}
{Marchesini} D.,  {van Dokkum} P.~G.,  {Forster Schreiber} N.~M.,  {Franx} M.,
  {Labbe'} I.,    {Wuyts} S.,  2008, ArXiv e-prints

\bibitem[\protect\citeauthoryear{{Martin}, {Small}, {Schiminovich}, {Wyder},
  {P{\'e}rez-Gonz{\'a}lez}, {Johnson}, {Wolf} \& {Barlow}}{{Martin}
  et~al.}{2007}]{Martin07}
{Martin} D.~C.,  {Small} T.,  {Schiminovich} D.,  {Wyder} T.~K.,
  {P{\'e}rez-Gonz{\'a}lez} P.~G.,  {Johnson} B.,  {Wolf} C.,    {Barlow} T.~A.
  e.~a.,  2007, \apjs, 173, 415

\bibitem[\protect\citeauthoryear{{Matteucci}}{{Matteucci}}{1994}]{Matteucci94}
{Matteucci} F.,  1994, \aap, 288, 57

\bibitem[\protect\citeauthoryear{{Menci}, {Fiore}, {Perola} \&
  {Cavaliere}}{{Menci} et~al.}{2004}]{Menci04}
{Menci} N.,  {Fiore} F.,  {Perola} G.~C.,    {Cavaliere} A.,  2004, \apj, 606,
  58

\bibitem[\protect\citeauthoryear{{Menci}, {Fiore}, {Puccetti} \&
  {Cavaliere}}{{Menci} et~al.}{2008}]{Menci08}
{Menci} N.,  {Fiore} F.,  {Puccetti} S.,    {Cavaliere} A.,  2008, \apj, 686,
  219

\bibitem[\protect\citeauthoryear{{Mobasher}, {Dahlen}, {Hopkins}, {Scoville},
  {Capak}, {Rich}, {Sanders}, {Schinnerer}, {Ilbert}, {Salvato} \&
  {Sheth}}{{Mobasher} et~al.}{2009}]{Mobasher09}
{Mobasher} B.,  {Dahlen} T.,  {Hopkins} A.,  {Scoville} N.~Z.,  {Capak} P.,
  {Rich} R.~M.,  {Sanders} D.~B.,  {Schinnerer} E.,  {Ilbert} O.,  {Salvato}
  M.,    {Sheth} K.,  2009, \apj, 690, 1074

\bibitem[\protect\citeauthoryear{{Monaco}, {Fontanot} \& {Taffoni}}{{Monaco}
  et~al.}{2007}]{Monaco07}
{Monaco} P.,  {Fontanot} F.,    {Taffoni} G.,  2007, \mnras, 375, 1189

\bibitem[\protect\citeauthoryear{{Monaco}, {Murante}, {Borgani} \&
  {Fontanot}}{{Monaco} et~al.}{2006}]{Monaco06}
{Monaco} P.,  {Murante} G.,  {Borgani} S.,    {Fontanot} F.,  2006, \apjl, 652,
  L89

\bibitem[\protect\citeauthoryear{{Monaco}, {Salucci} \& {Danese}}{{Monaco}
  et~al.}{2000}]{Monaco00}
{Monaco} P.,  {Salucci} P.,    {Danese} L.,  2000, \mnras, 311, 279

\bibitem[\protect\citeauthoryear{{Monaco}, {Theuns}, {Taffoni}, {Governato},
  {Quinn} \& {Stadel}}{{Monaco} et~al.}{2002}]{Monaco02}
{Monaco} P.,  {Theuns} T.,  {Taffoni} G.,  {Governato} F.,  {Quinn} T.,
  {Stadel} J.,  2002, \apj, 564, 8

\bibitem[\protect\citeauthoryear{{Nagashima}, {Lacey}, {Okamoto}, {Baugh},
  {Frenk} \& {Cole}}{{Nagashima} et~al.}{2005}]{Nagashima05}
{Nagashima} M.,  {Lacey} C.~G.,  {Okamoto} T.,  {Baugh} C.~M.,  {Frenk} C.~S.,
    {Cole} S.,  2005, \mnras, 363, L31

\bibitem[\protect\citeauthoryear{{Navarro}, {Frenk} \& {White}}{{Navarro}
  et~al.}{1996}]{Navarro96}
{Navarro} J.~F.,  {Frenk} C.~S.,    {White} S.~D.~M.,  1996, \apj, 462, 563

\bibitem[\protect\citeauthoryear{{Neistein}, {van den Bosch} \&
  {Dekel}}{{Neistein} et~al.}{2006}]{Neistein06}
{Neistein} E.,  {van den Bosch} F.~C.,    {Dekel} A.,  2006, \mnras, 372, 933

\bibitem[\protect\citeauthoryear{{Nelan}, {Smith}, {Hudson}, {Wegner}, {Lucey},
  {Moore}, {Quinney} \& {Suntzeff}}{{Nelan} et~al.}{2005}]{Nelan05}
{Nelan} J.~E.,  {Smith} R.~J.,  {Hudson} M.~J.,  {Wegner} G.~A.,  {Lucey}
  J.~R.,  {Moore} S.~A.~W.,  {Quinney} S.~J.,    {Suntzeff} N.~B.,  2005, \apj,
  632, 137

\bibitem[\protect\citeauthoryear{{Noeske}, {Faber}, {Weiner}, {Koo}, {Primack},
  {Dekel}, {Papovich}, {Conselice}, {Le Floc'h}, {Rieke}, {Coil}, {Lotz},
  {Somerville} \& {Bundy}}{{Noeske} et~al.}{2007}]{Noeske07}
{Noeske} K.~G.,  {Faber} S.~M.,  {Weiner} B.~J.,  {Koo} D.~C.,  {Primack}
  J.~R.,  {Dekel} A.,  {Papovich} C.,  {Conselice} C.~J.,  {Le Floc'h} E.,
  {Rieke} G.~H.,  {Coil} A.~L.,  {Lotz} J.~M.,  {Somerville} R.~S.,    {Bundy}
  K.,  2007, \apjl, 660, L47

\bibitem[\protect\citeauthoryear{{Pannella}, {Hopp}, {Saglia}, {Bender},
  {Drory}, {Salvato}, {Gabasch} \& {Feulner}}{{Pannella}
  et~al.}{2006}]{Pannella06}
{Pannella} M.,  {Hopp} U.,  {Saglia} R.~P.,  {Bender} R.,  {Drory} N.,
  {Salvato} M.,  {Gabasch} A.,    {Feulner} G.,  2006, \apjl, 639, L1

\bibitem[\protect\citeauthoryear{{Panter}, {Jimenez}, {Heavens} \&
  {Charlot}}{{Panter} et~al.}{2007}]{Panter07}
{Panter} B.,  {Jimenez} R.,  {Heavens} A.~F.,    {Charlot} S.,  2007, \mnras,
  378, 1550

\bibitem[\protect\citeauthoryear{{Papovich}, {Cool}, {Eisenstein}, {Le Floc'h},
  {Fan}, {Kennicutt} Jr., {Smith}, {Rieke} \& {Vestergaard}}{{Papovich}
  et~al.}{2006}]{Papovich06}
{Papovich} C.,  {Cool} R.,  {Eisenstein} D.,  {Le Floc'h} E.,  {Fan} X.,
  {Kennicutt} Jr. R.~C.,  {Smith} J.~D.~T.,  {Rieke} G.~H.,    {Vestergaard}
  M.,  2006, \aj, 132, 231

\bibitem[\protect\citeauthoryear{{P{\'e}rez-Gonz{\'a}lez}, {Rieke}, {Villar},
  {Barro}, {Blaylock}, {Egami}, {Gallego}, {Gil de Paz}, {Pascual}, {Zamorano}
  \& {Donley}}{{P{\'e}rez-Gonz{\'a}lez} et~al.}{2008}]{PerezGonzalez08}
{P{\'e}rez-Gonz{\'a}lez} P.~G.,  {Rieke} G.~H.,  {Villar} V.,  {Barro} G.,
  {Blaylock} M.,  {Egami} E.,  {Gallego} J.,  {Gil de Paz} A.,  {Pascual} S.,
  {Zamorano} J.,    {Donley} J.~L.,  2008, \apj, 675, 234

\bibitem[\protect\citeauthoryear{{Pipino}, {Devriendt}, {Thomas}, {Silk} \&
  {Kaviraj}}{{Pipino} et~al.}{2008}]{Pipino08}
{Pipino} A.,  {Devriendt} J.~E.~G.,  {Thomas} D.,  {Silk} J.,    {Kaviraj} S.,
  2008, ArXiv e-prints

\bibitem[\protect\citeauthoryear{{Pozzetti}, {Bolzonella}, {Lamareille},
  {Zamorani}, {Franzetti}, {Le F{\`e}vre}, {Iovino} \& {Temporin}}{{Pozzetti}
  et~al.}{2007}]{Pozzetti07}
{Pozzetti} L.,  {Bolzonella} M.,  {Lamareille} F.,  {Zamorani} G.,  {Franzetti}
  P.,  {Le F{\`e}vre} O.,  {Iovino} A.,    {Temporin} S. e.~a.,  2007, \aap,
  474, 443

\bibitem[\protect\citeauthoryear{{Salim}, {Rich}, {Charlot}, {Brinchmann},
  {Johnson}, {Schiminovich}, {Seibert} \& {Mallery}}{{Salim}
  et~al.}{2007}]{Salim07}
{Salim} S.,  {Rich} R.~M.,  {Charlot} S.,  {Brinchmann} J.,  {Johnson} B.~D.,
  {Schiminovich} D.,  {Seibert} M.,    {Mallery} R. e.~a.,  2007, \apjs, 173,
  267

\bibitem[\protect\citeauthoryear{{Santini}, {Fontana}, {Grazian}, {Salimbeni},
  {Fiore}, {Fontanot}, {Boutsia}, {Castellano}, {Cristiani}, {De Santis},
  {Gallozzi}, {Giallongo}, {Menci}, {Nonino}, {Paris}, {Pentericci} \&
  {Vanzella}}{{Santini} et~al.}{2009}]{Santini09}
{Santini} P.,  {Fontana} A.,  {Grazian} A.,  {Salimbeni} S.,  {Fiore} F.,
  {Fontanot} F.,  {Boutsia} K.,  {Castellano} M.,  {Cristiani} S.,  {De Santis}
  C.,  {Gallozzi} S.,  {Giallongo} E.,  {Menci} N.,  {Nonino} M.,  {Paris} D.,
  {Pentericci} L.,    {Vanzella} E.,  2009, ArXiv e-prints

\bibitem[\protect\citeauthoryear{{Savaglio}, {Glazebrook}, {Le Borgne},
  {Juneau}, {Abraham}, {Chen}, {Crampton}, {McCarthy}, {Carlberg}, {Marzke},
  {Roth}, {J{\o}rgensen} \& {Murowinski}}{{Savaglio} et~al.}{2005}]{Savaglio05}
{Savaglio} S.,  {Glazebrook} K.,  {Le Borgne} D.,  {Juneau} S.,  {Abraham}
  R.~G.,  {Chen} H.-W.,  {Crampton} D.,  {McCarthy} P.~J.,  {Carlberg} R.~G.,
  {Marzke} R.~O.,  {Roth} K.,  {J{\o}rgensen} I.,    {Murowinski} R.,  2005,
  \apj, 635, 260

\bibitem[\protect\citeauthoryear{{Schiminovich}, {Wyder}, {Martin}, {Johnson},
  {Salim}, {Seibert}, {Treyer} \& {Budav{\'a}ri}}{{Schiminovich}
  et~al.}{2007}]{Schiminovich07}
{Schiminovich} D.,  {Wyder} T.~K.,  {Martin} D.~C.,  {Johnson} B.~D.,  {Salim}
  S.,  {Seibert} M.,  {Treyer} M.~A.,    {Budav{\'a}ri} T. e.~a.,  2007, \apjs,
  173, 315

\bibitem[\protect\citeauthoryear{{Sheth} \& {Tormen}}{{Sheth} \&
  {Tormen}}{1999}]{ShethTormen99}
{Sheth} R.~K.,  {Tormen} G.,  1999, \mnras, 308, 119

\bibitem[\protect\citeauthoryear{{Somerville}, {Hopkins}, {Cox}, {Robertson} \&
  {Hernquist}}{{Somerville} et~al.}{2008}]{Somerville08}
{Somerville} R.~S.,  {Hopkins} P.~F.,  {Cox} T.~J.,  {Robertson} B.~E.,
  {Hernquist} L.,  2008, \mnras, 391, 481

\bibitem[\protect\citeauthoryear{{Somerville} \& {Kolatt}}{{Somerville} \&
  {Kolatt}}{1999}]{SomervilleKolatt99}
{Somerville} R.~S.,  {Kolatt} T.~S.,  1999, \mnras, 305, 1

\bibitem[\protect\citeauthoryear{{Somerville}, {Primack} \&
  {Faber}}{{Somerville} et~al.}{2001}]{Somerville01}
{Somerville} R.~S.,  {Primack} J.~R.,    {Faber} S.~M.,  2001, \mnras, 320, 504

\bibitem[\protect\citeauthoryear{{Springel}, {White}, {Jenkins}, {Frenk},
  {Yoshida}, {Gao}, {Navarro}, {Thacker}, {Croton}, {Helly}, {Peacock}, {Cole},
  {Thomas}, {Couchman}, {Evrard}, {Colberg} \& {Pearce}}{{Springel}
  et~al.}{2005}]{Springel05}
{Springel} V.,  {White} S.~D.~M.,  {Jenkins} A.,  {Frenk} C.~S.,  {Yoshida} N.,
   {Gao} L.,  {Navarro} J.,  {Thacker} R.,  {Croton} D.,  {Helly} J.,
  {Peacock} J.~A.,  {Cole} S.,  {Thomas} P.,  {Couchman} H.,  {Evrard} A.,
  {Colberg} J.,    {Pearce} F.,  2005, \nat, 435, 629

\bibitem[\protect\citeauthoryear{{Stringer}, {Benson}, {Bundy}, {Ellis} \&
  {Quetin}}{{Stringer} et~al.}{2009}]{Stringer09}
{Stringer} M.~J.,  {Benson} A.~J.,  {Bundy} K.,  {Ellis} R.~S.,    {Quetin}
  E.~L.,  2009, \mnras, 393, 1127

\bibitem[\protect\citeauthoryear{{Taffoni}, {Mayer}, {Colpi} \&
  {Governato}}{{Taffoni} et~al.}{2003}]{Taffoni03}
{Taffoni} G.,  {Mayer} L.,  {Colpi} M.,    {Governato} F.,  2003, \mnras, 341,
  434

\bibitem[\protect\citeauthoryear{{Thomas}}{{Thomas}}{1999}]{Thomas99}
{Thomas} D.,  1999, \mnras, 306, 655

\bibitem[\protect\citeauthoryear{{Thomas}, {Maraston}, {Bender} \& {Mendes de
  Oliveira}}{{Thomas} et~al.}{2005}]{Thomas05}
{Thomas} D.,  {Maraston} C.,  {Bender} R.,    {Mendes de Oliveira} C.,  2005,
  \apj, 621, 673

\bibitem[\protect\citeauthoryear{{Tonini}, {Maraston}, {Devriendt}, {Thomas} \&
  {Silk}}{{Tonini} et~al.}{2008}]{Tonini08}
{Tonini} C.,  {Maraston} C.,  {Devriendt} J.,  {Thomas} D.,    {Silk} J.,
  2008, ArXiv e-prints

\bibitem[\protect\citeauthoryear{{Trager}, {Faber} \& {Dressler}}{{Trager}
  et~al.}{2008}]{Trager08}
{Trager} S.~C.,  {Faber} S.~M.,    {Dressler} A.,  2008, \mnras, 386, 715

\bibitem[\protect\citeauthoryear{{Trager}, {Faber}, {Worthey} \&
  {Gonz{\'a}lez}}{{Trager} et~al.}{2000a}]{Trager00b}
{Trager} S.~C.,  {Faber} S.~M.,  {Worthey} G.,    {Gonz{\'a}lez} J.~J.,  2000a,
  \aj, 120, 165

\bibitem[\protect\citeauthoryear{{Trager}, {Faber}, {Worthey} \&
  {Gonz{\'a}lez}}{{Trager} et~al.}{2000b}]{Trager00a}
{Trager} S.~C.,  {Faber} S.~M.,  {Worthey} G.,    {Gonz{\'a}lez} J.~J.,  2000b,
  \aj, 119, 1645

\bibitem[\protect\citeauthoryear{{Trager} \& {Somerville}}{{Trager} \&
  {Somerville}}{2008}]{TragerSomerville08}
{Trager} S.~C.,  {Somerville} R.~S.,  2008, \mnras, accepted

\bibitem[\protect\citeauthoryear{{Ueda}, {Akiyama}, {Ohta} \& {Miyaji}}{{Ueda}
  et~al.}{2003}]{Ueda03}
{Ueda} Y.,  {Akiyama} M.,  {Ohta} K.,    {Miyaji} T.,  2003, \apj, 598, 886

\bibitem[\protect\citeauthoryear{{van den Bosch}, {Aquino}, {Yang}, {Mo},
  {Pasquali}, {McIntosh}, {Weinmann} \& {Kang}}{{van den Bosch}
  et~al.}{2008}]{vdBosch08}
{van den Bosch} F.~C.,  {Aquino} D.,  {Yang} X.,  {Mo} H.~J.,  {Pasquali} A.,
  {McIntosh} D.~H.,  {Weinmann} S.~M.,    {Kang} X.,  2008, \mnras, 387, 79

\bibitem[\protect\citeauthoryear{{Vergani}, {Scodeggio}, {Pozzetti}, {Iovino},
  {Franzetti}, {Garilli}, {Zamorani} \& {Maccagni}}{{Vergani}
  et~al.}{2008}]{Vergani08}
{Vergani} D.,  {Scodeggio} M.,  {Pozzetti} L.,  {Iovino} A.,  {Franzetti} P.,
  {Garilli} B.,  {Zamorani} G.,    {Maccagni} D. e.~a.,  2008, \aap, 487, 89

\bibitem[\protect\citeauthoryear{{Viola}, {Monaco}, {Borgani}, {Murante} \&
  {Tornatore}}{{Viola} et~al.}{2008}]{Viola08}
{Viola} M.,  {Monaco} P.,  {Borgani} S.,  {Murante} G.,    {Tornatore} L.,
  2008, \mnras, 383, 777

\bibitem[\protect\citeauthoryear{{Wang}, {De Lucia}, {Kitzbichler} \&
  {White}}{{Wang} et~al.}{2008}]{Wang08}
{Wang} J.,  {De Lucia} G.,  {Kitzbichler} M.~G.,    {White} S.~D.~M.,  2008,
  \mnras, 384, 1301

\bibitem[\protect\citeauthoryear{{Weinmann}, {van den Bosch}, {Yang}, {Mo},
  {Croton} \& {Moore}}{{Weinmann} et~al.}{2006}]{Weinmann06b}
{Weinmann} S.~M.,  {van den Bosch} F.~C.,  {Yang} X.,  {Mo} H.~J.,  {Croton}
  D.~J.,    {Moore} B.,  2006, \mnras, 372, 1161

\bibitem[\protect\citeauthoryear{{White} \& {Frenk}}{{White} \&
  {Frenk}}{1991}]{White91}
{White} S.~D.~M.,  {Frenk} C.~S.,  1991, \apj, 379, 52

\bibitem[\protect\citeauthoryear{{Worthey}}{{Worthey}}{1994}]{Worthey94}
{Worthey} G.,  1994, \apjs, 95, 107

\bibitem[\protect\citeauthoryear{{Worthey}, {Faber} \& {Gonzalez}}{{Worthey}
  et~al.}{1992}]{Worthey92}
{Worthey} G.,  {Faber} S.~M.,    {Gonzalez} J.~J.,  1992, \apj, 398, 69

\bibitem[\protect\citeauthoryear{{Zheng}, {Bell}, {Papovich}, {Wolf},
  {Meisenheimer}, {Rix}, {Rieke} \& {Somerville}}{{Zheng}
  et~al.}{2007}]{Zheng07}
{Zheng} X.~Z.,  {Bell} E.~F.,  {Papovich} C.,  {Wolf} C.,  {Meisenheimer} K.,
  {Rix} H.-W.,  {Rieke} G.~H.,    {Somerville} R.,  2007, \apjl, 661, L41

\end{thebibliography}

\label{lastpage}

\end{document}